\documentclass[10pt,aps,prd,showpacs,superscriptaddress,notitlepage,longbibliography,nofootinbib]{revtex4-1}
\usepackage[usenames,dvipsnames]{color}
\usepackage{amsmath}
\usepackage{amssymb}
\usepackage{latexsym}
\usepackage{enumerate}
\usepackage{bbm}
\usepackage{amsthm}
\usepackage[colorlinks]{hyperref}
\usepackage{graphicx}
\usepackage{dcolumn}
\hypersetup{citecolor=red, linkcolor=blue}
\usepackage{subfigure}
\usepackage{footnote}
\usepackage{ulem}
\usepackage{mathrsfs}
\usepackage{accents}
\DeclareSymbolFontAlphabet{\mathrsfs}{rsfs}

\DeclareFontFamily{OT1}{rsfs}{} \DeclareFontShape{OT1}{rsfs}{m}{n}{<-7> rsfs5 <7-10> rsfs7 <10-> rsfs10}{}
\DeclareMathAlphabet{\mycal}{OT1}{rsfs}{m}{n}
\newcommand{\const}{\mathrm{const.}}
\newcommand{\interior}[1]{\accentset{\smash{\raisebox{0ex}{$\scriptstyle\circ$}}}{#1}\rule{0pt}{2.3ex}}
\newcommand{\instar}[1]{\accentset{\smash{\raisebox{0ex}{$\scriptstyle\star$}}}{#1}\rule{0pt}{2.3ex}}
\renewcommand{\emph}[1]{\textit{#1}}

\begin{document}

\title{Construction of high precision numerical single and binary black hole initial data}

\author{Georgios Doulis}
\email{gdoulis@phys.uoa.gr}
\affiliation{Institute of Theoretical Physics, Faculty of Physics, University of Warsaw}

\date{\today} 

\vspace{1cm}

\begin{abstract}

We present a novel implicit numerical implementation of the parabolic-hyperbolic 
formulation of the constraints of general relativity. The proposed method 
is unconditionally stable, has the advantage of not requiring the imposition 
of any boundary conditions in the strong field regime, and offers a holistic 
(all inclusive) approach to the construction of single and binary black hole 
initial data. The new implicit solver is extensively tested against known 
exact black hole solutions and is used to construct initial data for several 
single and binary black hole configurations.  
  
\end{abstract}

\maketitle

\numberwithin{equation}{section}


\section{Introduction}
\label{sec:intro}

With the observation of the signal GW151226 \cite{GW151226}, general 
relativity officially entered the long-awaited era of observation of 
gravitational waves. This signal together with several others that 
have been observed to date \cite{GWcatalog2018} constitute the cornerstone 
of the emerging field of gravitational wave astronomy. The observed 
signals contain invaluable information about the physical properties 
of the merging binaries and the process of merging itself. The extraction 
of this information calls for the development of analytical and/or 
numerical methods that can reproduce the observed waveforms to the 
highest degree of accuracy. Due to the highly non-linear nature of the 
inspiral and merger of the binary systems that emitted the observed 
signals, the use of numerical methods is indispensable. During the 
last decade, following the seminal work of Pretorius \cite{Pretorius2005}, 
more and more sophisticated numerical models have been developed to simulate 
the dynamics of binary systems of massive compact objects. The accuracy 
of all these simulations crucially depends on their initialisation. 
It is expected that any failures or errors involved in the construction 
of the initial data sets will not only affect their subsequent temporal 
development but also the conclusions we can draw from it. Therefore, 
the construction of initial data for binary systems of massive astrophysical 
objects that are free from all sorts of defects---of analytical or 
numerical nature---is of paramount importance for numerical relativity. 
Especially today due to the key role of numerical relativity in simulating 
the gravitational waveforms emitted from binary systems of black holes 
or neutron stars.

In their entirety, the binary black hole initial data that have been 
constructed to date with any of the existing numerical methods contain 
some amount of ``junk radiation", i.e. pre-existing high frequency 
gravitational radiation that is not produced by the investigated 
binaries. Unfortunately, all the efforts that have been made to resolve 
this issue by modifying and/or refining the existing elliptic methods 
have not been successful. The presence of ``junk radiation" in the 
initial data contaminates more or less, through their numerical 
evolution, the produced gravitational waveforms. Although the presence 
of ``junk radiation" does not seem to affect the stability and convergence 
of the numerical codes, it affects to some extent the accuracy with 
which the parameters characterising the binary systems are estimated. 
(Recall that the parameters of a binary system---e.g. the mass, spin, 
etc.---are chosen in such a way that the numerically computed waveforms 
agree to a given precision with the observed gravitational wave signals.) 

The origin of the ``junk radiation" has been attributed to specific 
assumptions---i.e. maximal slicing condition, conformal flatness---made 
during the construction of the initial data with any of the available 
methods, see e.g. \cite{Alcubierre2008}. Moreover, recent work 
\cite{Chu2014}, relates also its presence to the use of excision. 
According to \cite{Chu2014}, the use of excision to place the inner 
boundary of the computational domain inside the event horizon distorts 
the tidal interactions between the black holes and leads to the creation 
of ``junk radiation". The results in \cite{Chu2014} not only show a 
clear correlation between the use of excision and the presence of 
``junk radiation", but also indicate that any approach that avoids 
the imposition of boundary conditions in the strong field regime could 
very well lead to a significant reduction of the `junk radiation" in 
the initial data. The above observations demonstrate the necessity of 
developing alternative methods of constructing initial data that are 
not bounded to the use of the above assumptions and techniques. 

Recently, a completely novel formulation of Einstein's constraint 
equations that fulfils all the aforementioned requirements was put 
forward by R\'acz \cite{Racz2015}. In the context of this formulation, 
a $2+1$ decomposition of the constraints is performed in a way that 
they can be viewed as an evolutionary system: parabolic-hyperbolic 
or algebraic-hyperbolic. Accordingly, one of the spatial coordinates, 
let's say $z$, plays the role of the 
temporal coordinate along which the initial data, prescribed on the 
2-dimensional surface composed by the remaining two spatial coordinates, 
are evolved. Exactly this feature is that disengages us from the 
``junk radiation"-generating assumptions used in the various variations 
of the elliptic York-Lichnerowicz method. Moreover, in this setting 
the use of excision is not required even in the case we are slicing 
through the event horizon of a black hole, see Fig.~\ref{Fig:IBVP} 
for more details. These extremely appealing features of the $2+1$ 
formulation of the constraints make the construction of initial data 
with significantly reduced or entirely suppressed ``junk radiation" 
highly doable. 

The use of the assumptions of maximal slicing, conformal flatness, and 
purely longitudinal Bowen-York extrinsic curvature in the construction 
of the initial data is not only responsible for the generation of ``junk 
radiation" in these data, but also restricts considerably the spectrum 
of the possible black hole initial data that can be constructed. For 
example the condition of conformal flatness excludes immediately the 
possibility of constructing any stationary Kerr type black hole initial 
data. R\'acz's method, on the other hand, is not bounded to the above 
assumptions and for this it can be used to generate any physically admissible 
single and binary black hole initial data. 

In addition to the above extremely pleasant features, the parabolic-hyperbolic 
formulation offers a holistic (all inclusive) approach to the construction 
of both single and binary black hole configurations. In other words, we 
do not need different methods to construct different types of black hole 
initial data. Currently, for example, the Bowen-York and the superposed 
Kerr-Schild data method are used to generate Schwarzschild and Kerr type 
initial data, respectively. The new approach accommodates in a straightforward 
way the initial data constructed with both the above methods.

Another very appealing feature of the parabolic-hyperbolic formulation 
is that the total ADM quantities of the system can be estimated before 
solving the constraints \cite{Racz2017}. In particular, the total ADM 
mass, centre of mass, linear and angular momenta of the black hole system 
under investigation can be given in terms of the input parameters. 

The present paper is organised as follows. Sec.~\ref{sec:parab-hyperb} 
contains a brief overview of the basic features of the parabolic-hyperbolic 
formulation of the constraints. The initial-boundary value problem for 
our system of equations is set up in Sec.~\ref{sec:IBVP}. Next, in Sec.~\ref{sec:numer_scheme} 
the novel implicit numerical scheme that is going to be used to solve 
the constraints is briefly discussed and in Sec.~\ref{sec:exact_sol} 
it is tested against known exact single and binary black hole solutions. 
Finally, in Sec.~\ref{sec:results} we present our numerical results 
concerning dynamical single and binary black hole configurations.


\section{Theoretical background}
\label{sec:background}

In this section, we briefly discuss the parabolic-hyperbolic formulation 
of the constraints pioneered in \cite{Racz2015} and set up the corresponding 
initial-boundary value problem \cite{Racz2018,Racz2018-SM}. 

\subsection{The constraints as a parabolic-hyperbolic system}
\label{sec:parab-hyperb} 

Vacuum initial data on a given three-dimensional Riemannian manifold 
$\Sigma$ are constructed by solving the vacuum constraints for the pair 
of symmetric tensors $(h_{ij}, K_{ij})$, where $h_{ij}$ is a Riemannian 
three-metric. In vacuum the Hamiltonian and momentum constraints read 
\cite{Choquet-Bruhat2009}: 
\begin{align}
 \label{hamiltonian}
 {}^{(3)} &R + K^2 -K_{ij} K^{ij} = 0, \\
 \label{momentum}
 & D_j K^j{}_i  - D_i K = 0,
\end{align} 
where $K = h^{ij}K_{ij}$, ${}^{(3)}R$ and $D_i$ denote the scalar curvature 
and the covariant derivative operator associated with the three-metric 
$h_{ij}$, respectively. 

In order to formulate \eqref{hamiltonian} and \eqref{momentum} as an 
parabolic-hyperbolic system we have first to $2+1$ decompose $\Sigma$.
Accordingly, we assume that the topology of $\Sigma$ allows a smooth 
foliation by a one-parameter family $\mycal{S}_\rho$ of two-surfaces 
that are the $\rho = \mathrm{const.}$ level surfaces of a smooth function 
$\rho: \Sigma \rightarrow \mathbb{R}$. Next, we choose a vector field 
$\rho^i$ on $\Sigma$ that satisfies the condition $\rho^i\, \partial_i 
\rho = 1$ and assume that its integral curves meet the $\mycal{S}_\rho$ 
leaves precisely ones. Consider then its orthogonal decomposition 
\begin{equation*}
 \rho^i = \widehat{N}\,\widehat{n}^i + \widehat{N}^i,
\end{equation*} 
where $\widehat{N}$ and $\widehat{N}^i$ are the lapse function and shift 
vector of $\rho^i$. The unit normal to the $\mycal{S}_\rho$ level surfaces 
reads then  
\begin{equation*}
 \widehat{n}^i = \widehat{N}^{-1} \left(\rho^i - \widehat{N}^i \right). 
\end{equation*}

Using $\widehat{n}^i$ one can further orthogonally decompose the physical 
quantities $h_{ij}$ and $K_{ij}$ as follows
\begin{equation}
 \label{h_K_decomp}
  \begin{aligned}
   &h_{ij} = \widehat \gamma_{ij} + \widehat  n_i\, \widehat n_j, \\
   K_{ij} = \boldsymbol\kappa& \,\widehat n_i\, \widehat n_j + \widehat n_i \, {\bf k}_j +
   \widehat n_j\, {\bf k}_i + {\bf K}_{ij},
  \end{aligned}
\end{equation}
where $\widehat \gamma_{ij}$ is the two-metric induced on the level surfaces 
$\mycal{S}_\rho$ and $\boldsymbol\kappa, {\bf k}_i, {\bf K}_{ij}$ are the 
scalar, vector, and tensorial projections of $K_{ij}$ on $\mycal{S}_\rho$. 
In the following, for reasons that will become soon apparent, see also 
\cite{Racz2014,Racz2015}, instead of ${\bf K}_{ij}$ we will use its trace 
${\bf K} = \widehat\gamma^{kl}\, {\bf K}_{kl}$ and trace-free $\interior{\bf 
K}_{ij}={\bf K}_{ij}-\tfrac1{2}\, \widehat\gamma_{ij}\,{\bf K}$ parts. 

With the aforedescribed $2+1$ decomposition the original physical quantities 
$(h_{ij}, K_{ij})$ have been replaced by the seven new fields $(\widehat N, 
\widehat N^i, \widehat \gamma_{ij}; \boldsymbol\kappa, {\bf k}{}_{i}, {\bf K}, 
\interior{\bf K}_{ij})$. Note that the original and the new variables have 
the same number of independent components, i.e twelve, and thus the former 
can be equivalently represented by the latter, and vice versa. 

The system \eqref{hamiltonian}-\eqref{momentum} is underdetermined as it 
provides only four equations for the twelve independent components of the 
septuple $(\widehat N, \widehat N^i, \widehat \gamma_{ij}; \boldsymbol\kappa, 
{\bf k}_{i}, {\bf K}, \interior{\bf K}_{ij})$. Thus, one has to choose for 
which fields to solve the constraints. It turns out that in the evolutionary 
formulation of the constraints only two possibilities are allowed \cite{Racz2015}: 
the so-called parabolic-hyperbolic and algebraic-hyperbolic system. In the 
present work, we will focus on the former. In the parabolic-hyperbolic case 
the three fields $(\widehat N, {\bf k}{}_{i}, {\bf K})$ are subject to the 
constraints \eqref{hamiltonian}-\eqref{momentum} whereas the remaining four 
fields $(\widehat N^i, \widehat \gamma_{ij}, \boldsymbol\kappa, \interior{\bf 
K}_{ij})$ are freely specifiable throughout $\Sigma$. 

In this setting, the Hamiltonian constraint \eqref{hamiltonian} reduces to 
a non-linear parabolic partial differential equation (PDE) for $\widehat N$: 
\begin{equation}
 \label{hamiltonian_N}
 \instar{K} \left[\partial_{\rho} \widehat N - \widehat N^l\, \widehat D_l 
  \widehat N \right] - \widehat N^2\, \widehat D^l \widehat D_l \widehat N - 
  \mathcal{A}\,\widehat N - \mathcal{B}\,\widehat N{}^{3} = 0, 
\end{equation}
where $\widehat D_i$ and $\widehat R$ denote the covariant derivative operator 
and the scalar curvature associated with the two-metric $\widehat \gamma_{ij}$, 
respectively; $\mathcal{A} = \partial_\rho \instar{K} - \widehat N^l\, \widehat 
D_l \instar{K} + \tfrac{1}{2}(\instar{K}{}^2 + \instar{K}_{kl}\, \instar{K}{}^{kl})$ 
with $\instar{K}_{ij} = \tfrac12\mycal{L}_\rho \widehat \gamma_{ij} - \widehat 
D_{(i}\widehat N_{j)}$ and $\instar{K}  = \widehat \gamma^{ij} \instar{K}_{ij}$; 
and $\mathcal{B} = -\tfrac12 \left(\widehat R + 2\, \boldsymbol\kappa\, {\bf K} + 
\tfrac12\,{\bf K}^2 - 2\,{\bf k}{}^{l}\,{\bf k}{}_{l} - \interior{\bf K}_{kl}\, 
\interior{\bf K}{}^{kl}\right)$. In \cite{Racz2015} it was shown that the 
Hamiltonian constraint \eqref{hamiltonian_N} is uniformly parabolic in those 
subsets of $\Sigma$ where $\instar{K}$ can be guaranteed to be strictly 
positive or negative. In this subsets $\rho$ plays the role of ``time" and 
$\rho^i$ becomes a ``time-evolution" vector field. Notice that as $\instar{K}$ 
depends exclusively on the freely specifiable fields $(\widehat N^i, \widehat 
\gamma_{ij})$, its sign (at least locally) is adjustable according to the 
details of the specific problem under consideration. 

In a similar fashion, the momentum constraint \eqref{momentum} reduces to 
three linear hyperbolic PDEs for the remaining two constrained fields $({\bf 
k}_{i}, {\bf K})$: 
\begin{align}
 \label{momentum_kd}
  & \mycal{L}_{\widehat n} {\bf k}_i - \tfrac12\,\widehat D_i\, {\bf K} - \widehat 
  D_i\, \boldsymbol\kappa + \widehat D^l\, \interior{\bf K}_{li} + \widehat N^{-1} 
  \instar{K}\,{\bf k}_{i}  + \left( \boldsymbol\kappa - \tfrac12\, {\bf K} \right)
  \dot{\widehat n}_i - \dot{\widehat n}^l\, \interior{\bf K}_{li} = 0, \\ 
 \label{momentumK} 
  & \mycal{L}_{\widehat n}{\bf K} - \widehat D^l\, {\bf k}_l - \widehat N^{-1} \instar{K}
  \left( \boldsymbol\kappa - \tfrac12\, {\bf K} \right) + \widehat N^{-1}\,\interior{\bf
K}_{kl}
  \instar{K}{}^{kl}  + 2\,\dot{\widehat n}^l\, {\bf k}_{l}  = 0,
\end{align}
where $\dot{\widehat n}_k={\widehat n}^l D_l{\widehat n}_k = -\widehat
D_k(\ln{\widehat N})$. 
It is worth noting that when \eqref{momentum_kd}-\eqref{momentumK} are expressed 
in terms of (local) coordinates adapted both to the foliation $\mycal{S}_\rho$ 
and the vector field $\rho^i$, then they can be written as a first order 
symmetric hyperbolic system, see \cite{Racz2015}. (Notice that the momentum 
constraint \eqref{momentum} would not have been brought into this pleasant 
form if we were working with ${\bf K}_{ij}$ instead of its trace ${\bf K}$ 
and trace-free $\interior{\bf K}_{ij}$ parts.) 

It also noteworthy that given the values of the constraints fields $(\widehat 
N, {\bf k}_{i}, {\bf K})$ on some ``initial" surface $\mycal{S}_0$, it was 
proven in \cite{Racz2015} that solutions to the non-linear system 
\eqref{hamiltonian_N}-\eqref{momentumK} exist (at least locally) in a neighbourhood 
of $\mycal{S}_0$ and that the fields $(h_{ij}, K_{ij})$ reconstructed from 
these solutions satisfy the full constraint system \eqref{hamiltonian}-\eqref{momentum}. 
Notice also that the PDEs \eqref{hamiltonian_N}-\eqref{momentumK} are coupled 
to each other. 


\subsection{The initial-boundary value problem}
\label{sec:IBVP}

\begin{figure}[thb]
 \centering 
  \hspace{-1cm} 
  \subfigure[\,Single black hole configuration.]{
   \label{Fig:singleIBVP}
   \includegraphics[height=6cm,width=5.5cm]{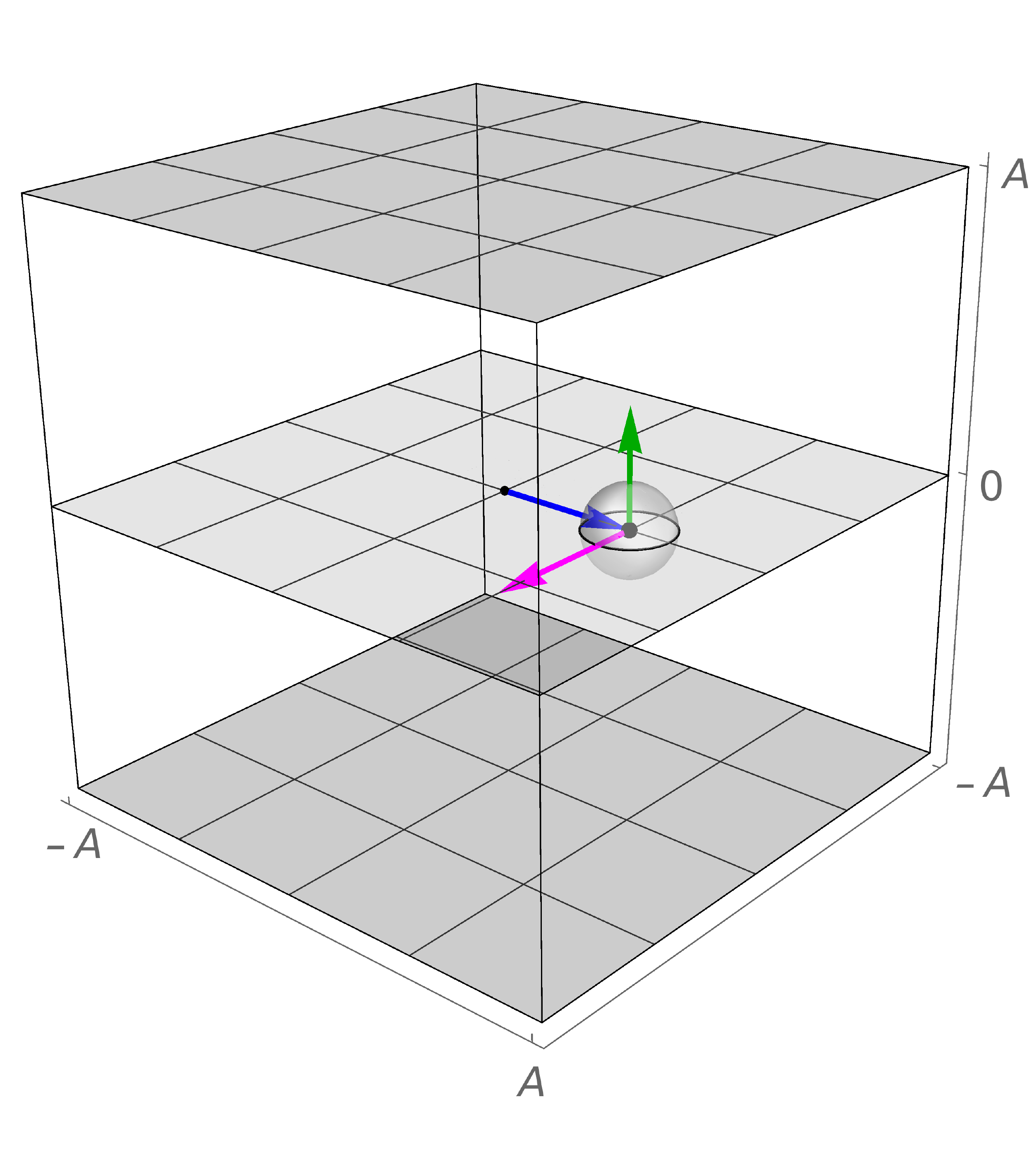}
   \put(-75,100){\scriptsize$\vec{d}$}
   \put(-58,108){\scriptsize$\vec{s}$}
   \put(-75,80){\scriptsize$\vec{\upsilon}$}
   \put(-33,33){$x$}
   \put(-122,31){$y$}
   \put(0,99){$z$}
  } \hspace{1cm} 
  \subfigure[\,Binary black hole configuration.]{
   \label{Fig:binaryIBVP}
   \includegraphics[height=6cm,width=5.5cm]{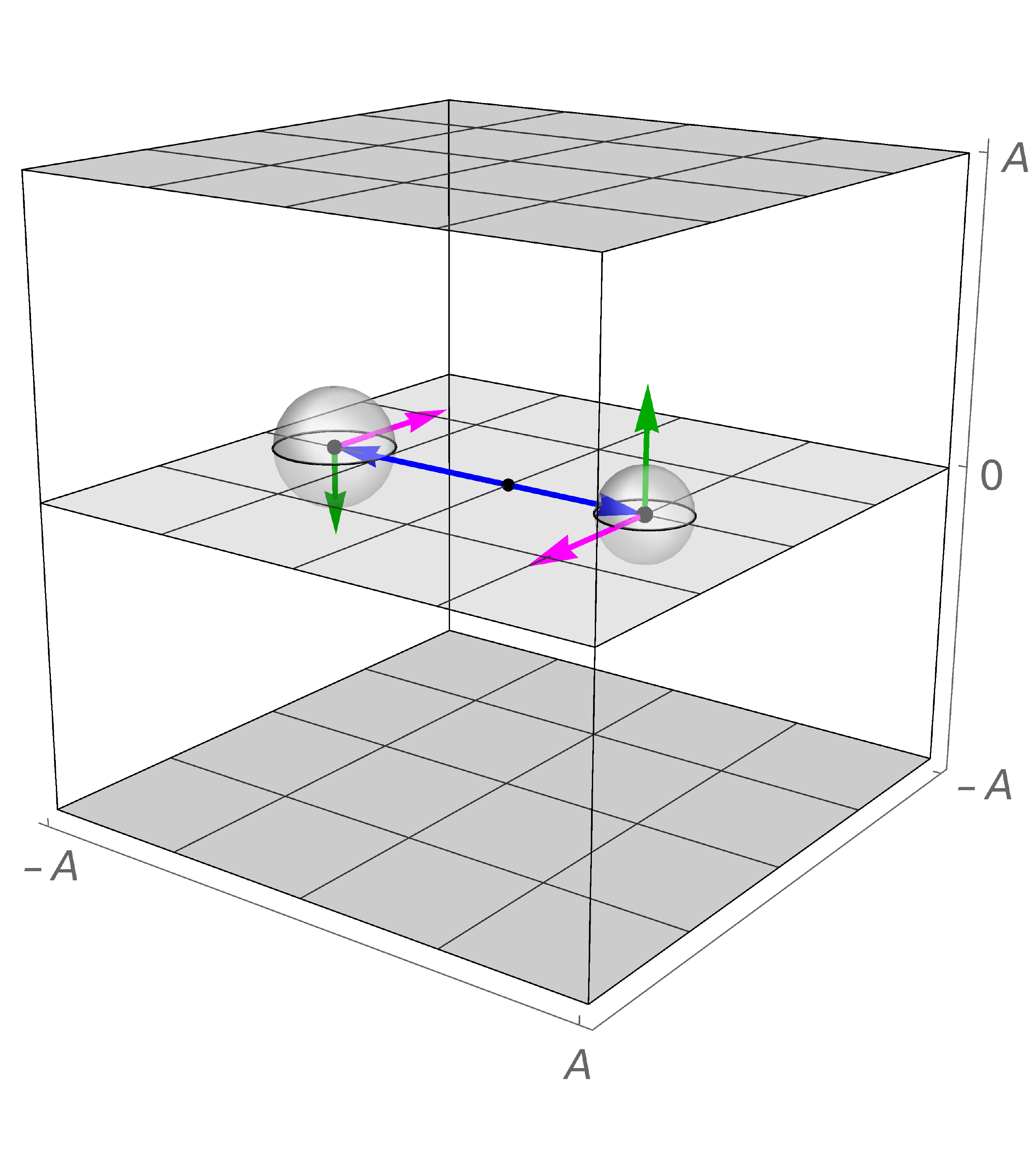}
   \put(-76,100){\scriptsize$\vec{d}^{[1]}$}
   \put(-57,110){\scriptsize$\vec{s}^{[1]}$}
   \put(-75,83){\scriptsize$\vec{\upsilon}^{[1]}$}
   \put(-100,95){\scriptsize$\vec{d}^{[2]}$}
   \put(-118,90){\scriptsize$\vec{s}^{[2]}$}
   \put(-96,112){\scriptsize$\vec{\upsilon}^{[2]}$}
   \put(-30,37){$x$}
   \put(-119,30){$y$}
   \put(0,100){$z$}
  } \hspace{-1cm}
  \caption{The initial data three-surface $\Sigma$. The initial data surface 
  is represented as a finite cube centered at the origin of $\mathbb{R}^3$ 
  with edges of length $2A$ for (a) single and (b) binary black hole configuration.}
 \label{Fig:IBVP} 
\end{figure}
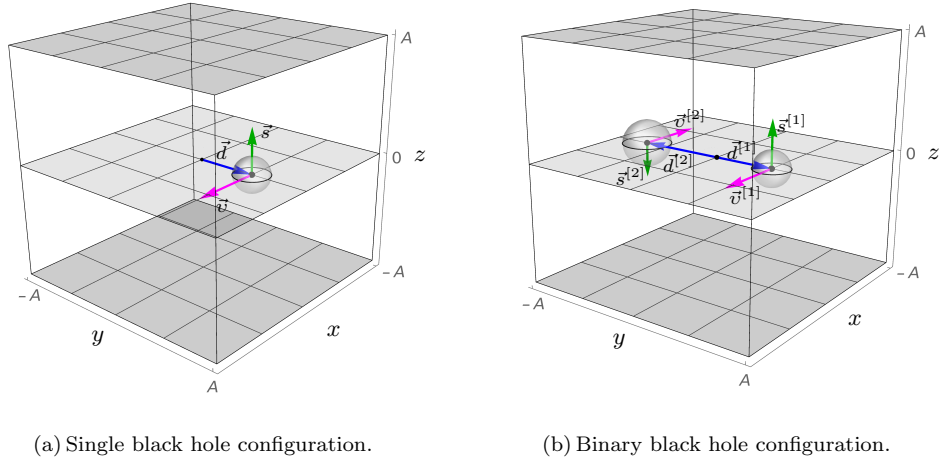

We set up now the initial-boundary value problem for the system 
\eqref{hamiltonian_N}-\eqref{momentumK}. The initial data three-surface 
$\Sigma$ is chosen to be a cube of finite side, centered at the origin of 
$\mathbb{R}^3$ as depicted in Fig.~\ref{Fig:IBVP}, with a boundary comprised 
of six squares with edges of length $2A$. By choosing the value of $A$ 
sufficiently large all the individual black holes involved in our constructions 
will be contained in this cubical domain with a sufficient margin.   

In the present work, we choose to work with Cartesian coordinates $(x,y,z)$ 
and to foliate $\Sigma$ with $z=\const$ two-surfaces. With these choices the 
constraints \eqref{hamiltonian_N}-\eqref{momentumK} for the fields $(\widehat 
N, {\bf k}_A, {\bf K})$ take respectively the form
\begin{align}
\label{hamiltonian_cart}
&\partial_z \widehat N = A_1\, \partial^2_{xx} \widehat N + A_2\, \partial^2_{yy} 
\widehat N + A_3\, \partial^2_{xy} \widehat N + A_4\, \partial_x \widehat N + A_5\, 
\partial_y \widehat N + \mathcal{A}\, \widehat N + \mathcal{B}\, \widehat N{}^3,\\
\label{momentum_kx_cart} 
&\partial_z {\bf k}_x = B_1\, \partial_x {\bf k}_x + B_2\, \partial_y {\bf k}_x +
B_3\, \partial_x {\bf K} + B_4,\\
\label{momentum_ky_cart} 
&\partial_z {\bf k}_y = C_1\, \partial_x {\bf k}_y + C_2\, \partial_y {\bf k}_y +
C_3\, \partial_y {\bf K} + C_4,\\ 
\label{momentum_K_cart}  
&\partial_z {\bf K} = D_1\, \partial_x {\bf K} + D_2\, \partial_y {\bf K} +
D_3\, \partial_x {\bf k}_x + D_4\, \partial_x {\bf k}_y + D_5\, \partial_y {\bf k}_x 
+ D_6\, \partial_y {\bf k}_y + D_7, 
\end{align}
where ${\bf k}_x, {\bf k}_y$ are the components of the vector ${\bf k}_A$ 
along the $x,y$ direction,\footnote{In the following, capital Latin indices 
will take the values $x$ and $y$.} respectively; the functions $\mathcal{A}$ 
and $\mathcal{B}$ have been defined above; and the exact expressions of the 
functions $A_i, B_i, C_i, D_i$ are given in Appendix~\ref{sec:appendixA}. In 
Sec.~\ref{sec:numer_implement} the above system will be solved numerically 
for several different black hole configurations. 

In the setting of Fig.~\ref{Fig:IBVP}, the considered black holes are placed 
on the $z=0$ plane and lie along the $y$-axis at distance $\vec{d}^{[i]}=d^{[i]} 
\hat{e}_y$ from the origin with linear velocity $\vec{\upsilon}^{[i]}=\upsilon^{[i]} 
\hat{e}_x$ along the $x$-direction and carry spin $\vec{s}^{[i]}=M^{[i]} a^{[i]} 
\hat{e}_z$ along the $z$-direction. (The unit vectors $\hat e_i$ are aligned 
along the respective positive directions.) In order to avoid configurations 
with extremal black holes or with naked singularities, in the following it 
will be always assumed that $|a^{[i]}| < M^{[i]}$.  

According to \cite{Racz2018,Racz2018-SM}, the values of the freely-specifiable 
fields $(\widehat N^A,\, \widehat \gamma_{AB}, \, \boldsymbol\kappa, \, 
\overset{\circ}{\rm\bf K}_{AB})$ must be prescribed throughout $\Sigma$ and 
for the remaining three constraint fields $(\widehat N,\, {\rm\bf k}{}_{A},\, 
{\bf K})$ the constraints \eqref{hamiltonian_cart}-\eqref{momentum_K_cart} 
must be solved. The latter requires the initialisation of the constraint fields 
on a $z=\const$ plane. One can prescribe initial data for the fields $(\widehat 
N,\, {\rm\bf k}{}_A,\, {\bf K})$ on one of the shaded (upper or lower) sides 
of the cubes in Fig.~\ref{Fig:IBVP} that are positioned at $z=\pm A$. Then, 
by solving the constraints \eqref{hamiltonian_cart}-\eqref{momentum_K_cart}, 
these data can be evolved along the $z$-streamlines towards the plane at $z=0$ 
where the black hole singularities are located. In addition, because of the 
finite nature of $\Sigma$, boundary conditions for the constrained fields must 
be also prescribed on the four vertical sides of $\Sigma$ located at $x=y=\pm A$. 

In order to provide the aforementioned values to the fields $(\widehat 
N,\, {\bf k}{}_{A},\, {\bf K}; \widehat N^A,\, \widehat \gamma_{AB}, \, 
\boldsymbol \kappa, \, \overset{\circ}{\rm\bf K}_{AB})$, we follow the 
proposal put forward in \cite{Racz2018,Racz2018-SM}. Therein,  Kerr-Schild 
type black hole data are used to provide the values of the constrained 
fields $(\widehat N,\, {\rm\bf k}{}_A,\, {\bf K})$ on the sides of $\Sigma$ 
and of the freely-specifiable fields $(\widehat N^A,\, \widehat \gamma_{AB},\, 
\boldsymbol\kappa,\, \overset{\circ}{\bf K}_{AB})$ throughout $\Sigma$. 
For an exhaustive and extremely detailed discussion of this topic see 
\cite{Racz2018,Racz2018-SM, Nakonieczna2017}. Here, we just present the 
form of the space-time metric for a generic boosted and displaced Kerr-Schild 
black hole
\begin{equation}
 \label{genericKS}
  g_{\alpha\beta} = \eta_{\alpha\beta} + 2\, H\, \ell_{\alpha}\, \ell_{\beta} 
\end{equation}
and for a  pair of superposed Kerr-Schild black holes
\begin{equation}
 \label{binaryKS}
  g_{\alpha\beta}= \eta_{\alpha\beta} + 2 H^{[1]}\, \ell_\alpha{}^{[1]} \, 
  \ell_\beta{}^{[1]} + 2 H^{[2]}\, \ell_\alpha{}^{[2]}\, \ell_\beta{}^{[2]}, 
\end{equation}
where, in inertial coordinates $(t,x,y,z)$ adapted to the background Minkowski 
metric $\eta_{\alpha\beta}$, we have 
\begin{equation*}
 H = \frac{r'^3 M}{r'^4 + {a^2 z^2}}, \quad 
 \ell_{\alpha} = \left(\gamma - \gamma\, \upsilon\, \ell'_1, - \gamma\, 
 \upsilon + \gamma\, \ell'_1, \ell'_2, \ell'_3 \right)
\end{equation*}
with $\ell'_{\alpha} = \left( 1, \frac{r'\,x'+a\,y'}{r'^2+a^2}, \frac{r'\,y'-
a\,x'}{r'^2+a^2}, \frac{z}{r'} \right)$ and  $r'^4 - (x'^2 + y'^2 + z^2 - 
a^2)\,r'^2 - a^2\,z^2 = 0$, $x' = \gamma\, x - \gamma\, \upsilon\,t, y' = y - 
d$, $\gamma = 1/\sqrt{1-\upsilon^2}$. 

Therefore, in the following, the values of the fields $(\widehat N^A,\, 
\widehat \gamma_{AB},\, \boldsymbol \kappa,\, \overset{\circ}{\bf K}_{AB})$ 
throughout $\Sigma$ and of $(\widehat N,\, {\bf k}_A,\, {\bf K})$ on the boundaries 
of $\Sigma$ for single and binary black hole configurations will be deduced 
from \eqref{genericKS} and \eqref{binaryKS}, respectively. A comment about 
the use of \eqref{binaryKS} is in order here. It turns out that although the
metric \eqref{binaryKS} does not satisfy Einstein's vacuum equations, it is 
a very good candidate for our purposes here as it is asymptotically flat and 
can be used to approximate the regions close to the individual black holes 
quite well \cite{Racz&Jeff2015}. At first sight, such a proposal seems at 
least preposterous. A closer look though puts things back in place again as 
the numerically computed solutions of the constraints 
\eqref{hamiltonian_cart}-\eqref{momentum_K_cart} will always differ from 
the corresponding fields that could be deduced analytically from the metric 
\eqref{binaryKS}.

\section{Numerical implementation}
\label{sec:numer_implement}

\subsection{Setting up the numerical scheme}
\label{sec:numer_scheme}

We choose to solve the constraints \eqref{hamiltonian_cart}-\eqref{momentum_K_cart} 
numerically using the so-called Alternating Direction Implicit (ADI) method. 
An implicit method was preferred over the explicit methods mainly because of 
its superior stability and convergence features in the treatment of the parabolic 
PDE \eqref{hamiltonian_cart}. It can be shown, see e.g. \cite{Douglas1955}, 
that for \eqref{hamiltonian_cart} the ADI method is unconditionally stable 
and second order in the evolutionary coordinate. The unconditional stability 
of the ADI method will enable us to get as close as possible to the $z=0$ 
plane where the black holes singularities are located. Moreover, the ADI 
method was preferred over the Crank-Nicolson implicit method because the use 
of the latter for two-dimensional parabolic PDEs like \eqref{hamiltonian_cart} 
becomes essentially impractical due to the high complexity involved. 

We use the method of lines to discretize the $1+2$ parabolic-hyperbolic 
system \eqref{hamiltonian_cart}-\eqref{momentum_K_cart}. Accordingly, the 
latter will be reduced to a system of ordinary differential equations by 
discretizing the spatial coordinates $(x,y)$ with finite difference techniques. 
In accordance with the setting of Sec.~\ref{sec:IBVP}, our two-dimensional 
computational domain is $(x,y) \in D = [-A, A]\times[-A, A]$, where $2A$ 
is the length of the edges of the cubes of Fig.~\ref{Fig:IBVP}. To obtain 
a finite representation of $D$ an equidistant two-dimensional grid 
$(x_i, y_j) = (-A + i\, h_x, -A + j\, h_y)$ with grid spacings 
$(h_x, h_y) = (2A/N_x, 2A/N_y)$ is introduced, where 
$i = 0, \dots, N_x$, $j = 0, \dots, N_y$ with $N_x+1, N_y+1$ being the number 
of grid-points along the $x,y$-direction, respectively. Notice that in the 
following, it will be always assumed that $N_x=N_y$, this is a requirement 
that is imposed by the ADI method. Each of the fields $\mathbb{F}_A = 
(\widehat N,\, \widehat N^i,\, \widehat \gamma_{ij},\, \boldsymbol\kappa,\, 
{\bf k}{}_{i},\, {\bf K},\, \overset{\circ}{\bf K}_{ij})$ is discretised in 
a similar fashion, i.e. $(\mathbb{F}_A)_{ij} = \mathbb{F}_A(x_i, y_j)$. 

Next, the spatial derivatives of the constraint fields are approximated 
with appropriate finite difference operators. As the ADI method is second 
order accurate in the spatial coordinates, we accordingly choose to use 
second order central difference operators to approximate the first and 
second derivatives appearing in \eqref{hamiltonian_cart}-\eqref{momentum_K_cart}.

The resulting semi-discrete system of ordinary differential equations will 
be evolved along the $z$-direction with the operator splitting ADI method 
introduced above. Remember that in the setting of Fig.~\ref{Fig:IBVP}, $z$ 
is the ``dynamical" coordinate along which evolution takes place. 

Now, in order to check the convergence of our numerical solutions, we define 
the convergence rate as follows  
\begin{equation}
 \label{conv_rate}
  \mathrm{CR} = \frac{\log_2(E_0/E_1)}{\log_2(h_0/h_1)},
\end{equation}
where $E_0$ and $E_1$ are the normalised $l^2$--norms of the errors of 
the numerical solutions $\mathcal{N}_0$ and $\mathcal{N}_1$ with resolution 
$h_0$ and $h_1$($<h_0$), respectively. The normalised $l^2$--norms $E_i$ 
will be computed with respect to the available exact solution $\mathcal{E}$, 
i.e. 
\[
E_i = \frac{1}{N+1} \sqrt{\sum_{k=1}^{N+1} \left| (\mathcal{N}_i)_k - \mathcal{E}_k \right|^2},
\]
or (when an exact solution is not available) with respect to the numerical 
simulation of the immediate higher resolution:\footnote{The author is grateful 
to I. R\'acz for pointing out this expression.}
\[
E_i = \frac{1}{N+1} \sqrt{\sum_{k=1}^{N+1} \left| (\mathcal{N}_i)_k - (\mathcal{N}_{i+1})_k \right|^2},
\]
where $N+1$ is the number of grid-points along the $x$ or the $y$ direction. 
(Recall that $N_x=N_y=N$.)

The code has been written from scratch in Python. A detailed description 
of the developed implicit numerical scheme and of the technical features 
of the resulting numerical algorithm will be publish elsewhere. 


\subsection{Testing the code with exact solutions}
\label{sec:exact_sol}

Before we start using our code to study numerically the parabolic-hyperbolic 
system \eqref{hamiltonian_cart}-\eqref{momentum_K_cart}, we will carry 
out---as one should always do---some numerical tests to check its performance 
in real-life scenarios. Known exact black hole solutions are the best 
candidates for this job. In the following, the developed numerical scheme 
will be tested in the single black hole case against the well-known Schwarzschild 
and Kerr solutions and in the binary case against the Brill-Lindquist solution. 


\subsubsection{Single black holes}
\label{sec:single_exact}

\noindent \textbf{Schwarzschild black hole}.
We start testing the performance of our code with a displaced boosted 
Schwarzschild black hole expressed in the analytical form \eqref{genericKS}. 
\begin{figure}[thb]
 \centering 
  \hspace{-0.2cm}
  \subfigure{
   \includegraphics[height=5.5cm,width=5cm]{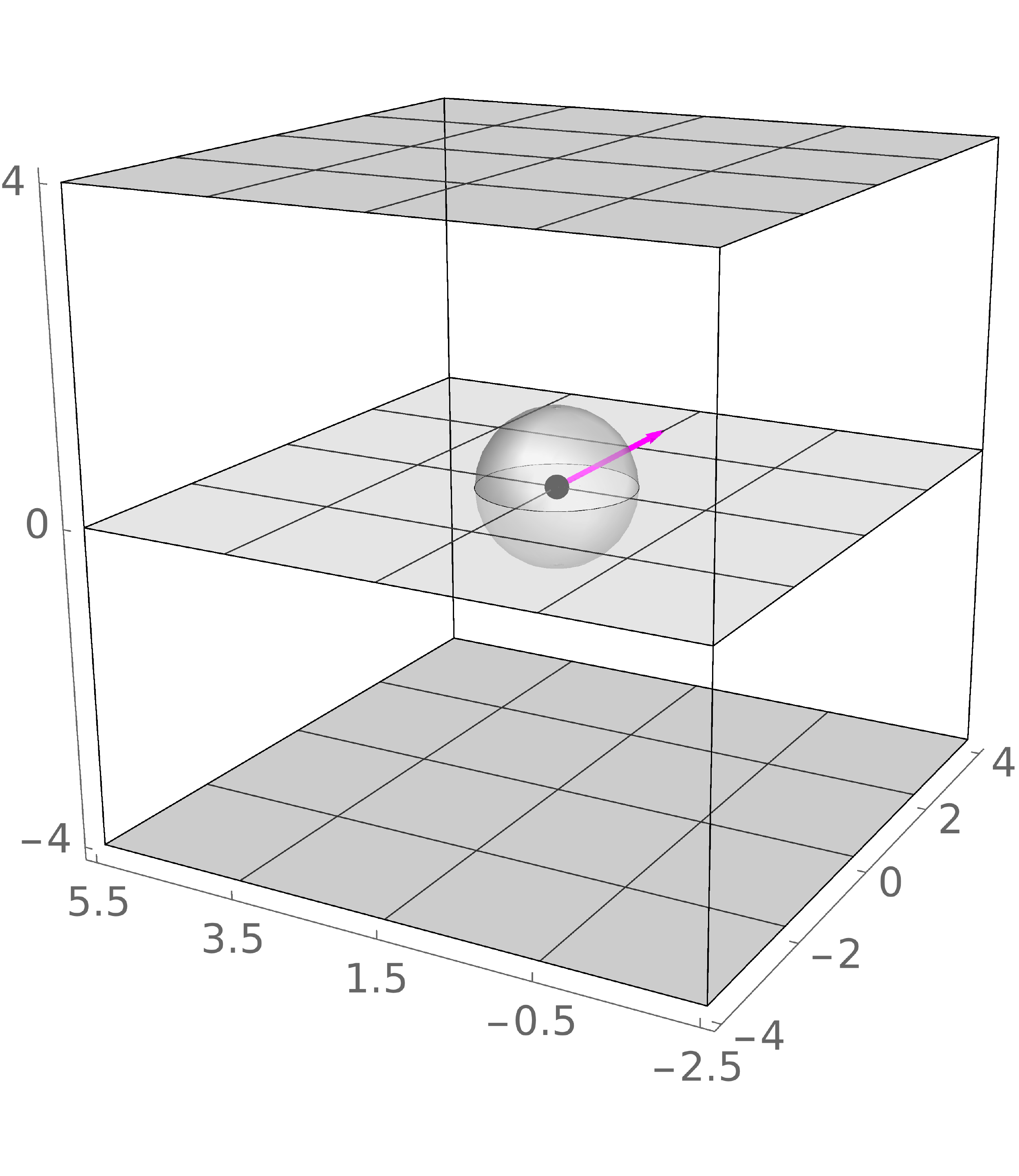}
  } \hspace{0.5cm}
  \subfigure{
   \includegraphics[height=5.5cm,width=8cm]{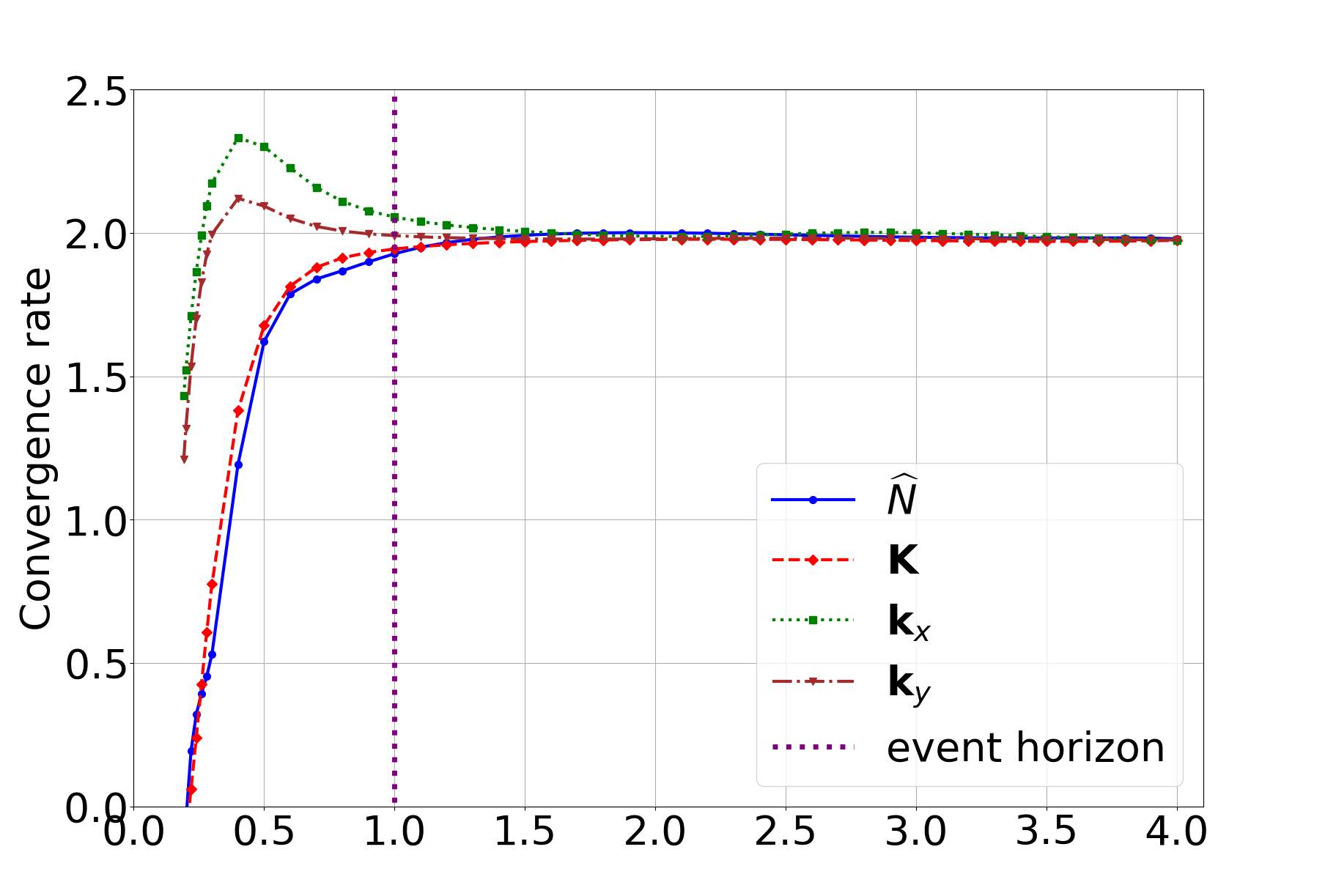}
  } \hspace{-1cm}
  \put(-290,100){\scriptsize$\vec{\upsilon}$}
  \put(-242,33){$x$}
  \put(-333,17){$y$}
  \put(-380,84){$z$}
  \put(-4,0){$z$}
  \caption{A displaced boosted Schwarzschild black hole. The left panel 
  illustrates the computational domain on which the constraints 
  \eqref{hamiltonian_cart}-\eqref{momentum_K_cart} are solved in the case 
  of a Schwarzschild black hole with input parameters $(M,\,d,\, \upsilon) 
  = (0.5,\,1.5,\,0.7)$. The dynamical behaviour of the convergence rate 
  \eqref{conv_rate} for each one of the constrained fields is depicted on 
  the right panel. (Notice the position of the event horizon at $z=1$.)}  
 \label{Fig:Schw_exact} 
\end{figure}
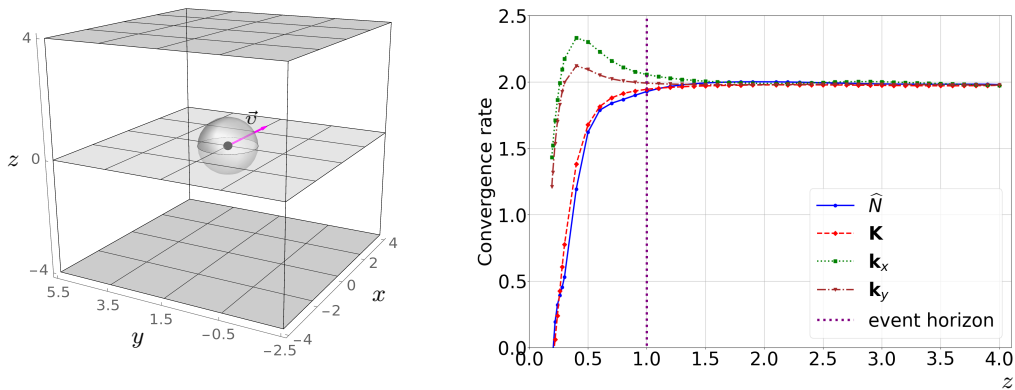
The input parameters characterising the considered black hole read $(M,\,
d,\,\upsilon) = (0.5,\,1.5,\,0.7)$. The left panel of Fig.~\ref{Fig:Schw_exact} 
depicts the location of the black hole, its event horizon, and the direction 
of its velocity on the computational domain $-4 \leq x \leq 4$,\, $-2.5 
\leq y \leq 5.5$,\, $0 \leq z \leq 4$ used in our numerical simulation. 
To initialise the $z$-evolution initial data for the constrained fields 
are prescribed on the upper shaded horizontal side of the cube located 
at $z = 4$. Subsequently, using the constraints 
\eqref{hamiltonian_cart}-\eqref{momentum_K_cart}, these initial data are 
evolved towards the $z=0$ plane where the black hole singularity is located. 

The resulting numerical solutions are highly accurate and our findings 
concerning their convergence properties are presented on the right panel 
of Fig.~\ref{Fig:Schw_exact}. Therein, the dynamical behaviour of the 
convergence rates for each one of the constrained fields is illustrated. 
It is clearly visible that during most of the evolution the convergence 
rates of all the constrained fields are around $2$, even when we are quite 
inside the event horizon. The expected drop of convergence while approaching 
the $z=0$ plane (where the singularity is located) can be considerable 
slowed down and delayed by increasing the resolution of our simulations. 
This quite challenging numerical feature is less alarming than one would 
expect, as it has been already shown in \cite{Nakonieczna2017}, that the 
accuracy of the produced data and the convergence properties of our numerical 
scheme close to the $z=0$ plane can be considerably improved by employing 
the so-called deviation form of the parabolic-hyperbolic system. 

\medskip

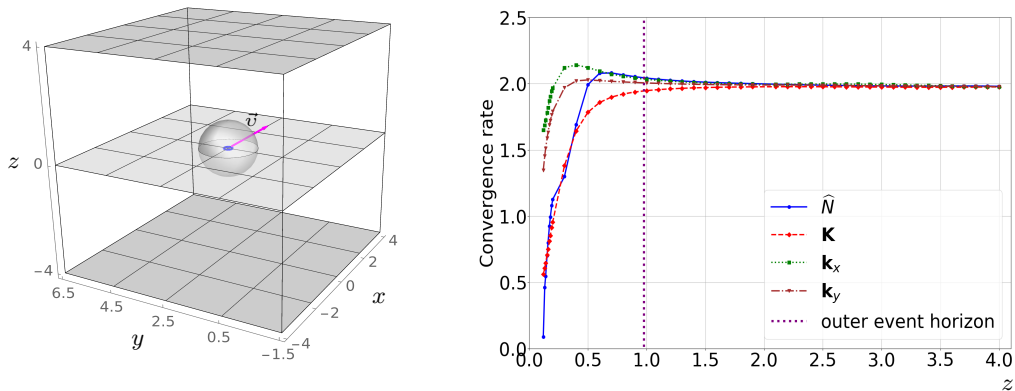
\begin{figure}[thb]
 \centering 
  \hspace{-0.2cm}
  \subfigure{
   \includegraphics[height=5.5cm,width=5.cm]{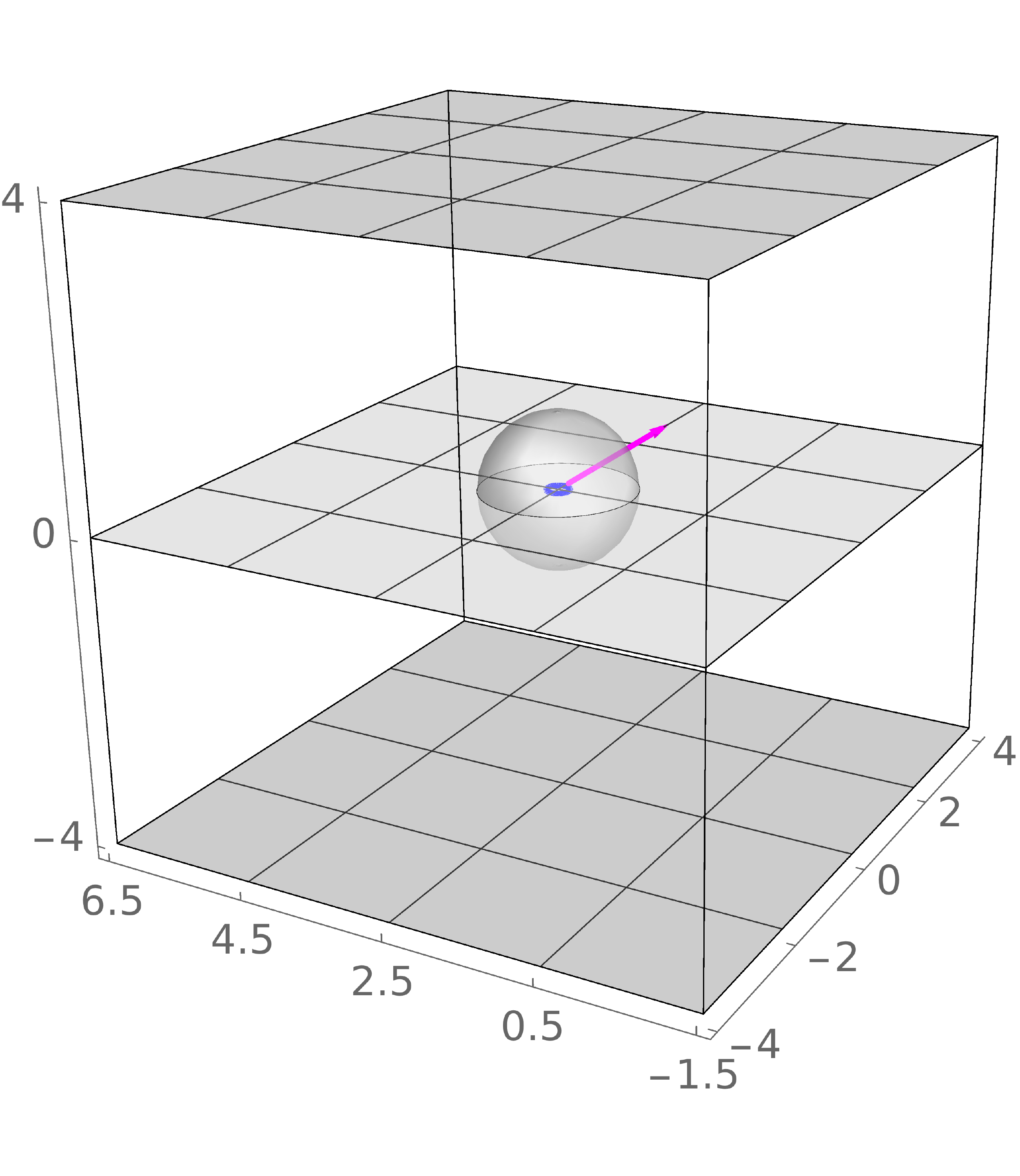}
  } \hspace{0.5cm}
  \subfigure{
   \includegraphics[height=5.5cm,width=8.cm]{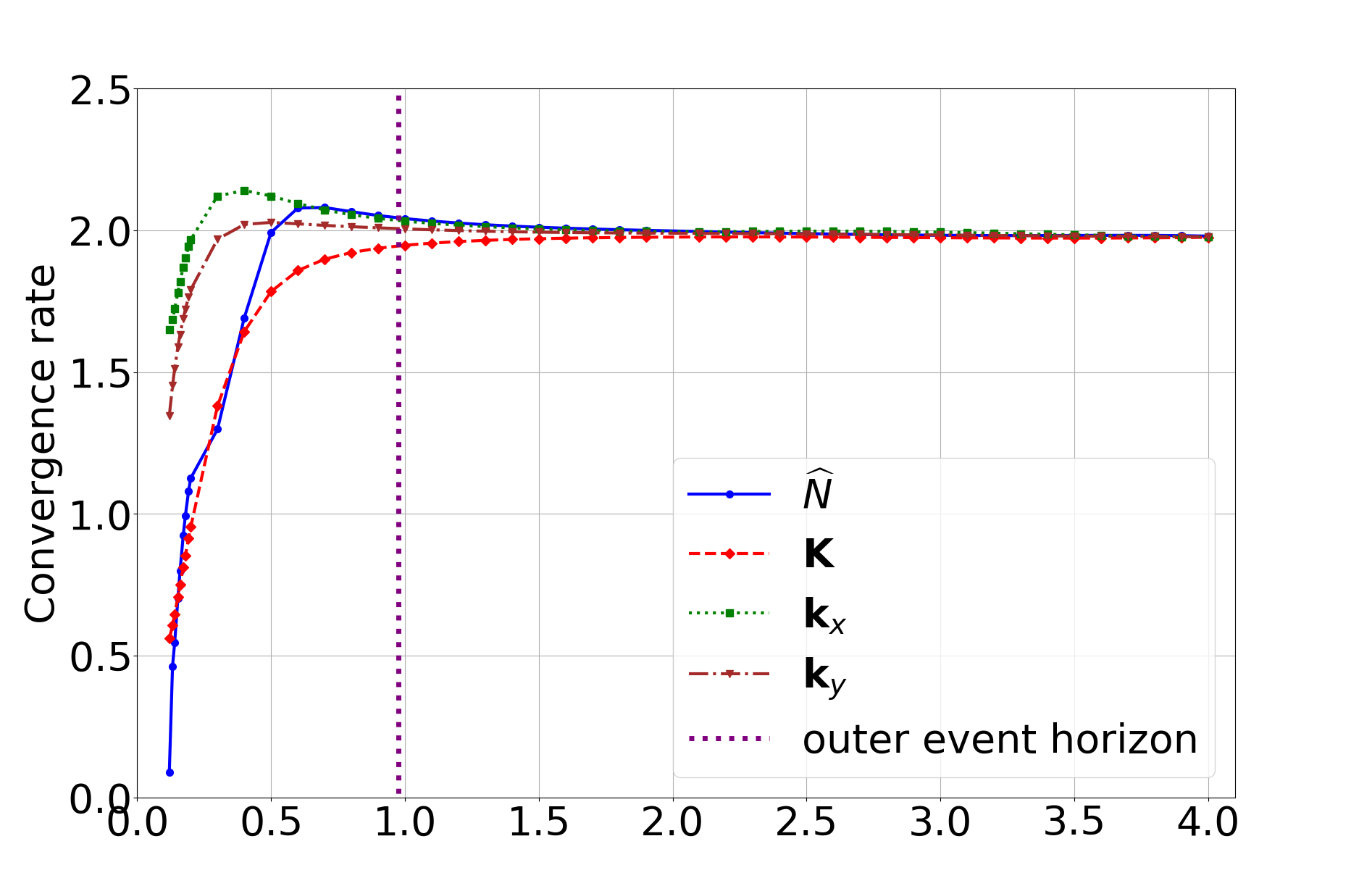}
  } \hspace{-1.cm}
  \put(-290,100){\scriptsize$\vec{\upsilon}$}
  \put(-242,33){$x$}
  \put(-333,17){$y$}
  \put(-380,84){$z$}
  \put(-4,0){$z$}
  \caption{A displaced boosted Kerr black hole. The left panel depicts 
  the computational domain on which initial data for a Kerr black hole 
  with input parameters $(M,\,d,\,\upsilon,\,a)=(0.5,\,2.5,\,0.6,\,0.15)$ 
  are constructed. Notice on the $z=0$ plane the characteristic ring 
  singularity of radius $r_{ring}=0.15$ related to the spin of the 
  black hole. The right panel illustrates the behaviour of the convergence 
  rates of the constrained fields during the $z$-evolution. (Notice 
  that the radius of the outer event horizon is $r_H \approx 0.977$.)}
 \label{Fig:Kerr_exact} 
\end{figure}

\noindent \textbf{Kerr black hole}. 
Next, we test our code with respect to a displaced boosted Kerr black 
hole given in the analytical form \eqref{genericKS} with input parameters 
$(M,\,d,\,\upsilon,\,a)=(0.5,\,2.5,\,0.6,\,0.15)$. Our numerical setting 
depicted on the left panel of Fig.~\ref{Fig:Kerr_exact} is quite similar 
to the one for the Schwarzschild black hole discussed above---this is the 
first instance of the holistic nature of the new approach. The location 
of the black hole, the outer event horizon, the direction of the velocity, 
and the ring singularity of radius $r_{ring}=0.15$---blue ring confined 
to the $z=0$ plane---are clearly visible on the left panel of Fig.~\ref{Fig:Kerr_exact}. 
The computational domain used is $-4 \leq x \leq 4$,\, $-1.5 \leq y \leq 
6.5$, and $0 \leq z \leq 4$.       

The dynamical behaviour of the convergence rate \eqref{conv_rate} for 
each one of the constrained fields $(\widehat N,\, {\bf k}_A,\, {\bf K})$ 
is illustrated on the right panel of Fig.~\ref{Fig:Kerr_exact}. The 
convergence rates have been computed with respect to the analytic 
expressions of the constrained fields deduced from the Kerr-Schild 
form \eqref{genericKS} of the considered Kerr black hole. As expected, 
during the evolution the convergence rates of all the constrained fields 
range around $2$ and drop gradually as we approach the $z=0$ plane.

%


\subsubsection{Brill-Lindquist binary black holes}
\label{sec:binary_exact}

The three-metric describing a pair of black holes at a moment of time 
symmetry is given in Cartesian coordinates by the expression
\begin{equation}
 \label{BL_exact}
 h_{ij}\, dx^i dx^j = \left(1 + \frac{1}{2}\sum_{i=1}^2\frac{M^{[i]}}{\left[x^2 +
 \left(y-d^{[i]}\right)^2 + z^2\right]} \right)^4 \left[dx^2 + dy^2 +  dz^2\right],
\end{equation}
where $M^{[i]}$ and $d^{[i]}$, with $i = 1,2$, denote the bare masses 
and the displacements from the origin, along the $y$-axis, of the two 
black holes, respectively. (Notice that the Brill-Lindquist metric cannot 
be written in Kerr-Schild form.) Due to the time-symmetric nature of the 
Brill-Lindquist data $K_{ij}$ vanishes, therefore all the fields, i.e. 
$(\boldsymbol \kappa,\, {\bf k}_A,\, {\bf K},\, \overset{\circ}{\bf 
K}_{AB})$, derived from $K_{ij}$ also vanish. Thus, we are left with 
only two freely-specifiable fields $(\widehat N^A,\, \widehat \gamma_{AB})$ 
and one constrained field $\widehat N$. Consequently, we have to solve 
only the parabolic equation \eqref{hamiltonian_cart} for $\widehat 
N$---the rest equations of the parabolic-hyperbolic system 
\eqref{hamiltonian_cart}-\eqref{momentum_K_cart} are trivially satisfied. 
The initial and boundary values for $\widehat N$ and the values of 
$(\widehat N^A,\, \widehat \gamma_{AB})$ throughout $\Sigma$ are deduced 
from \eqref{BL_exact}. 
\begin{figure}[thb]
 \centering 
  \hspace{-0.2cm}
  \subfigure{
   \includegraphics[height=5.5cm,width=5.cm]{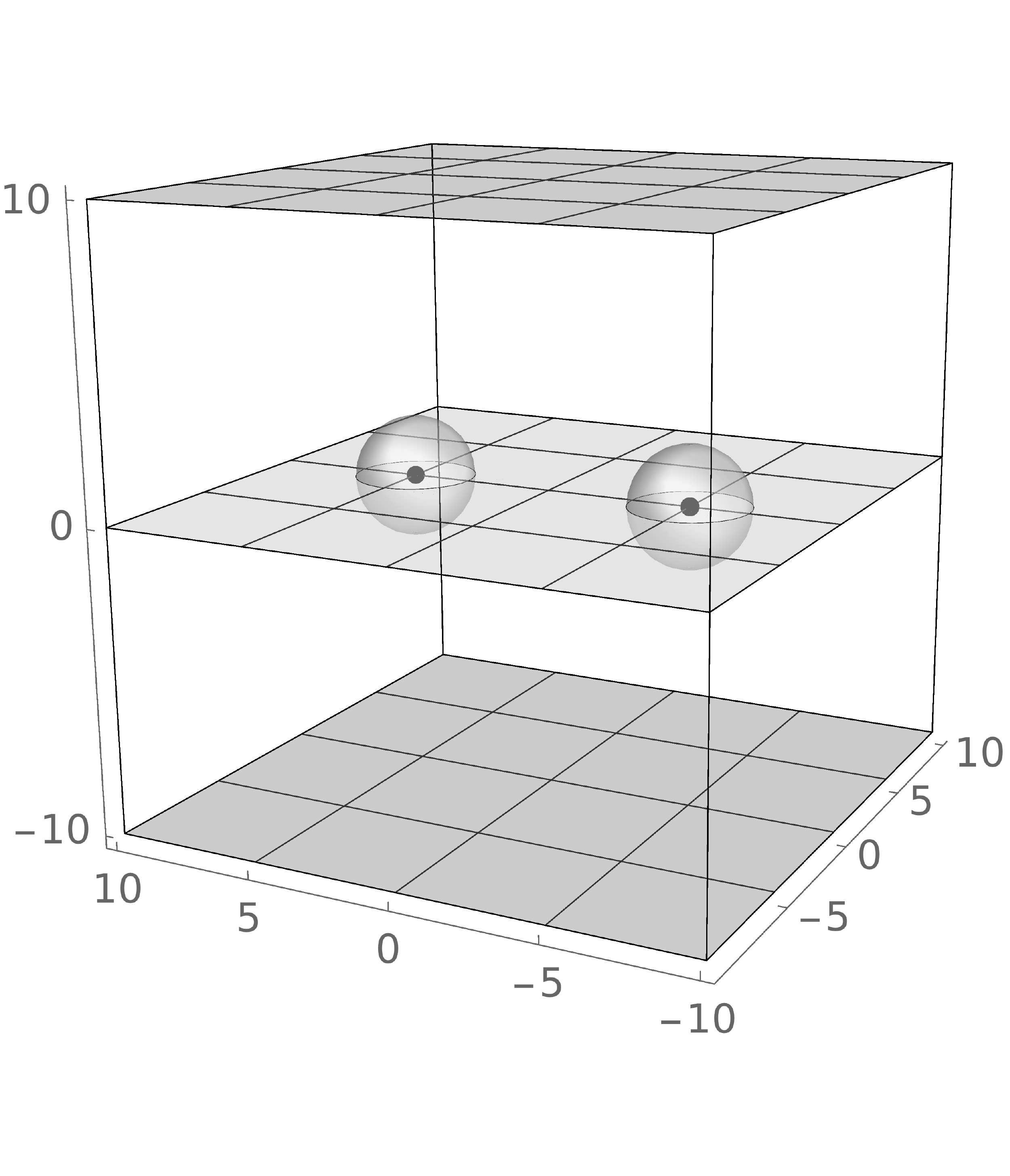}
  } \hspace{0.7cm}
  \subfigure{
   \includegraphics[height=5.5cm,width=8.cm]{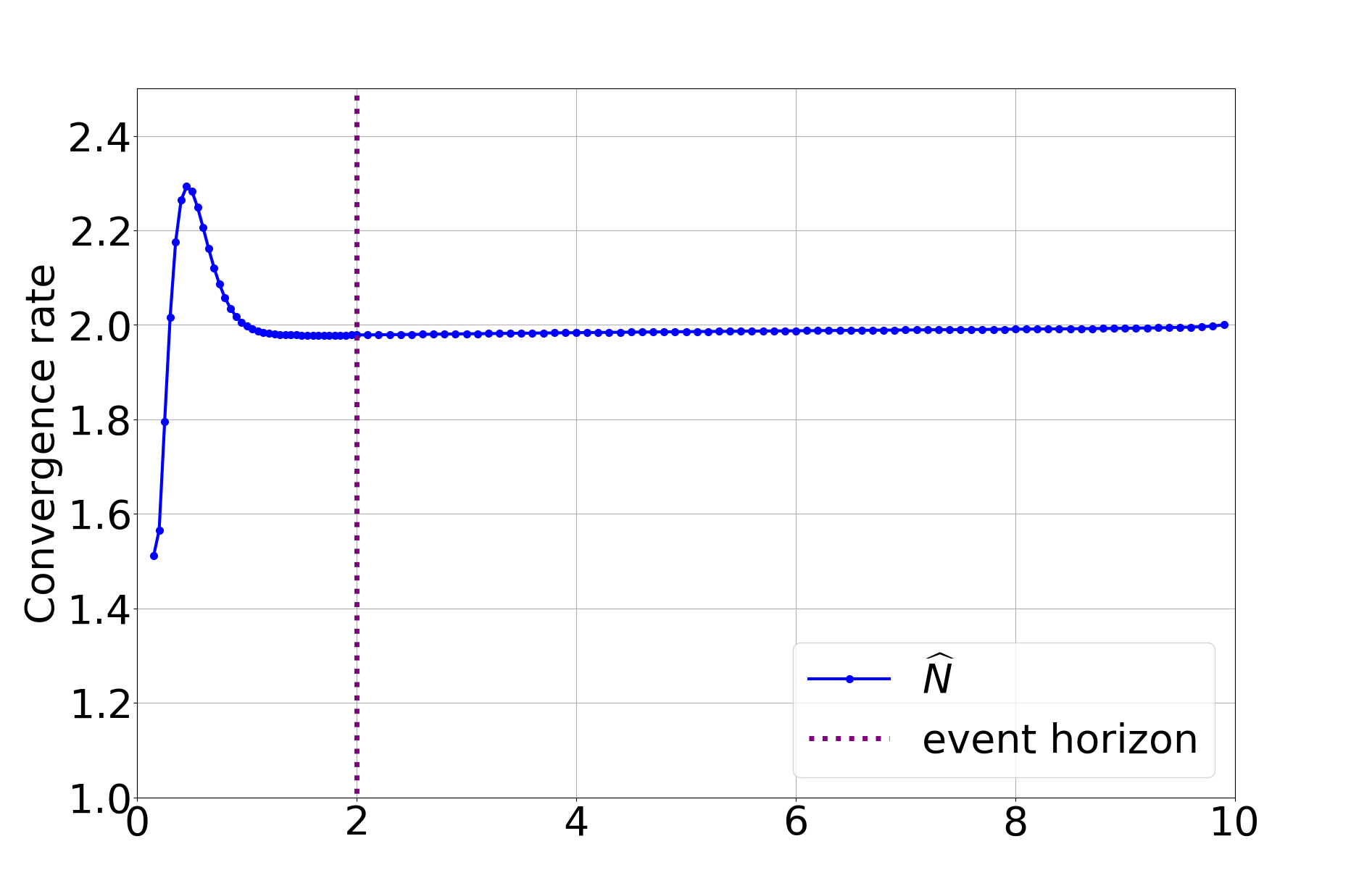}
  } \hspace{-1cm}
  \put(-248,37){$x$}
  \put(-335,20){$y$}
  \put(-383,84){$z$}
  \put(-4,0){$z$}
  \caption{Brill-Lindquist black holes. The two black holes of equal 
  mass $M^{[1]} = M^{[2]} = 1$ lie symmetrically to the origin on the 
  $y$-axis with $d^{[1]} = -d^{[2]} = 5$. The left panel also depicts 
  the computational domain used in solving the only non-trivially 
  satisfied constraint, i.e. the Hamiltonian constraint 
  \eqref{hamiltonian_cart}. The dynamical behaviour of the convergence 
  rate of the only non-trivial constraint field $\widehat{N}$ is depicted 
  on the right panel.}
 \label{Fig:BL_exact} 
\end{figure}

The left panel of Fig.~\ref{Fig:BL_exact} depicts our numerical arrangement. 
A pair of non-boosted black holes with masses $M^{[1]}=M^{[2]} = 1$ are 
positioned on the $y$-axis at distances $d^{[1]} = -d^{[2]} = 5$ from the 
origin. Both black holes are placed on the $z=0$ plane. Our computational 
domain is $-10 \leq x \leq10,\, -10 \leq y \leq10$, and $0 \leq z \leq10$. 
Initial data for $\widehat N$ are prescribed on the upper shaded horizontal 
side of the cube located at $z=10$ and are evolved with \eqref{hamiltonian_cart} 
towards the $z=0$ plane where the black hole singularities are positioned. 

The behaviour of the convergence rate \eqref{conv_rate} with $z$ of the 
only non-trivial constrained field $\widehat{N}$ is depicted on the right 
panel of Fig.~\ref{Fig:BL_exact}. The convergence rates during the evolution 
are around $2$, even on $z$-slices well inside the event horizon, and they 
only start dropping extremely close to the $z=0$ plane. 


\subsection{Numerical results: Dynamical black hole configurations}
\label{sec:results}

The results of the previous section constitute strong evidence that our 
code can successfully reproduce the exact single and binary black hole 
solutions considered in Sec.~\ref{sec:exact_sol}. Our findings confirm 
the expected convergence and stability features of the implicit method 
we are using. Therefore, we are confident enough to proceed further in 
the numerical investigation of the parabolic-hyperbolic form of the 
constraints and look for more general non-stationary solutions of the 
system \eqref{hamiltonian_cart}-\eqref{momentum_K_cart}. 

\subsubsection{Distorted Kerr}
\label{sec:dist_Kerr}

First, we look for dynamical single black hole solutions of the system 
\eqref{hamiltonian_cart}-\eqref{momentum_K_cart}. Specifically, we 
construct initial data for dynamical single black hole configurations 
resulting from the distortion of the stationary Kerr black hole considered 
in Sec.~\ref{sec:single_exact}.
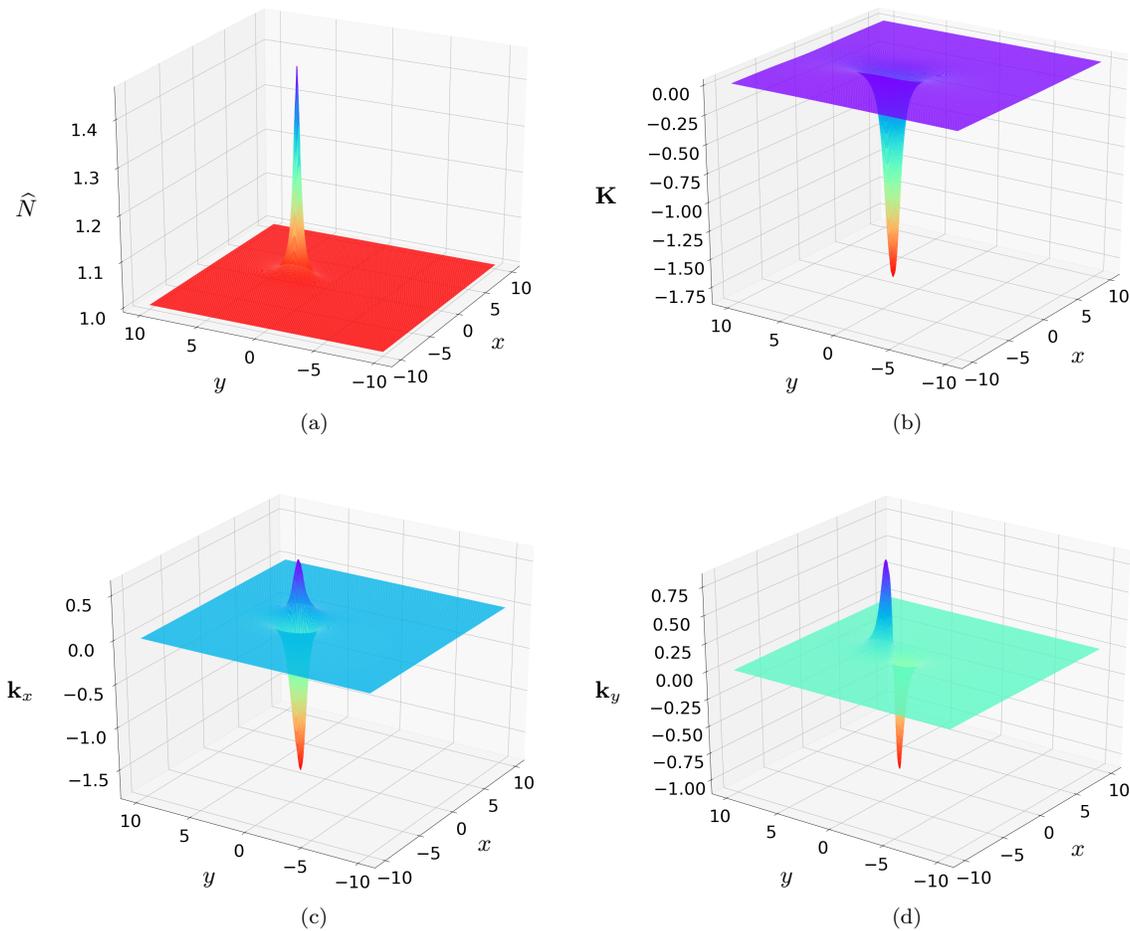
\begin{figure}[htb]
 \centering
  \subfigure[]{
   \includegraphics[scale=0.185]{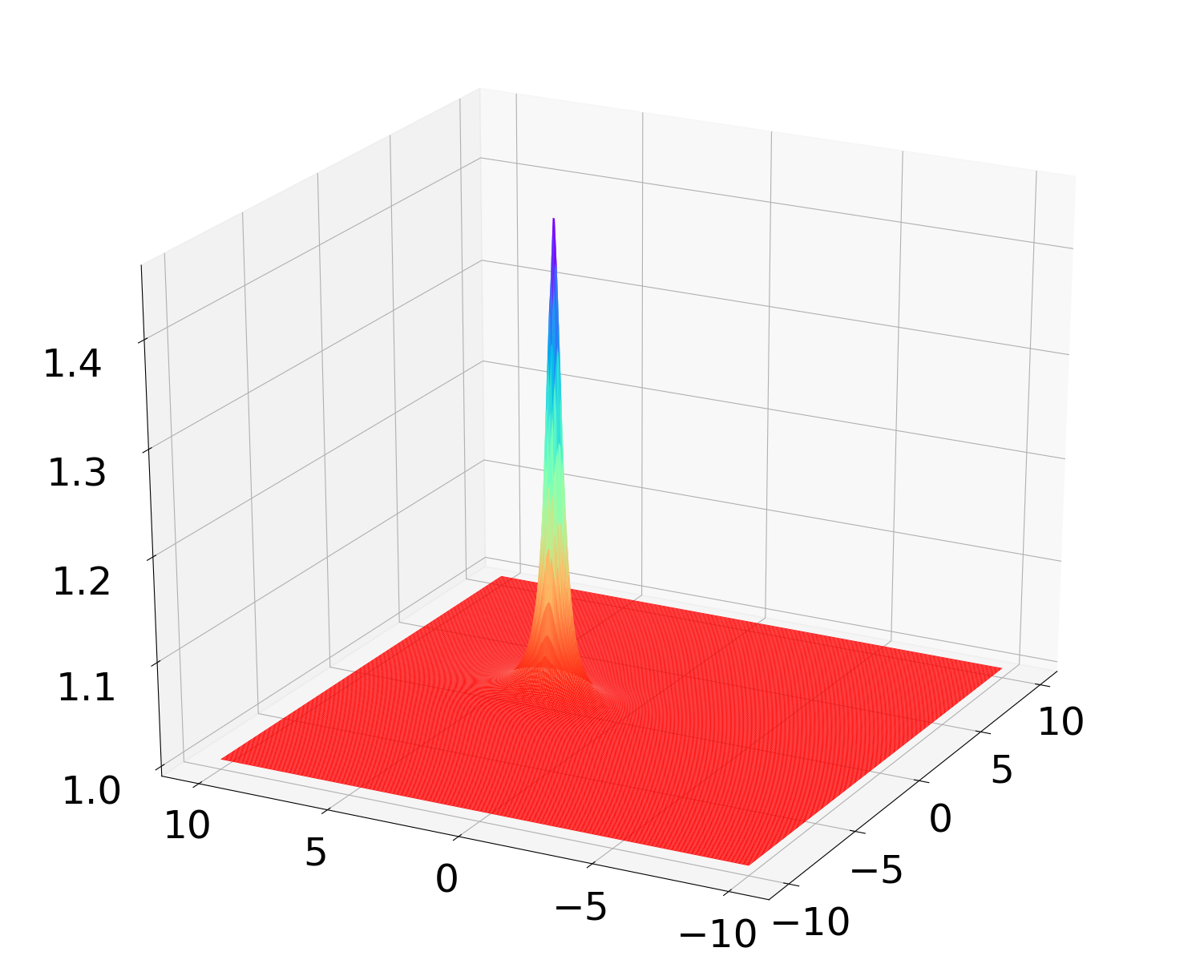}
   \put(-215,70){$\widehat{N}$} 
   \put(-35,20){$x$}
   \put(-140,5){$y$} 
  } \hspace{0.5cm}
  \subfigure[]{
   \includegraphics[scale=0.185]{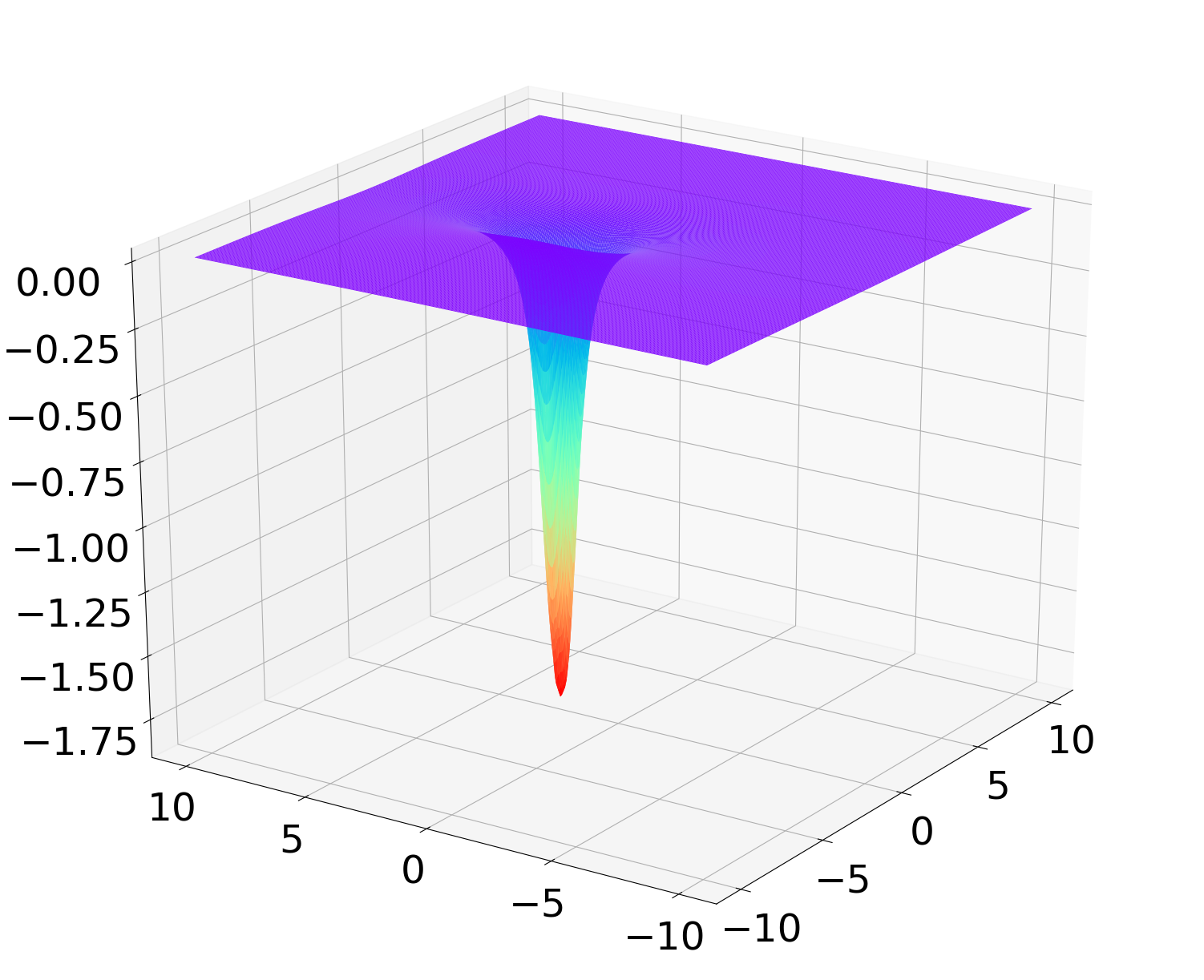}
   \put(-220,75){${\bf K}$} 
   \put(-40,15){$x$}
   \put(-148,5){$y$}    
  }
  \subfigure[]{
   \includegraphics[scale=0.185]{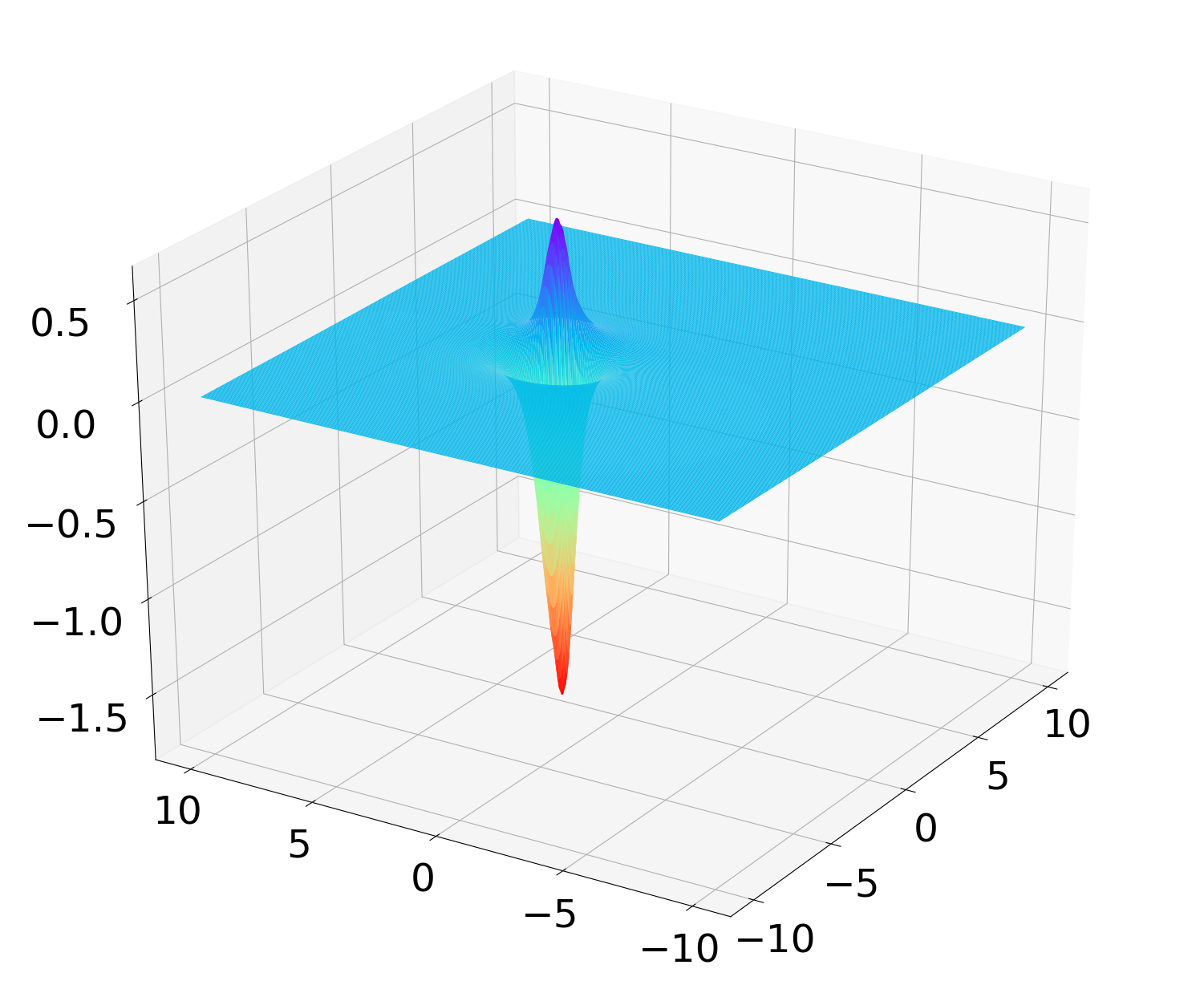}
   \put(-218,75){${\bf k}_x$}
   \put(-40,17){$x$}
   \put(-144,5){$y$}    
  } \hspace{0.5cm}
  \subfigure[]{
   \includegraphics[scale=0.185]{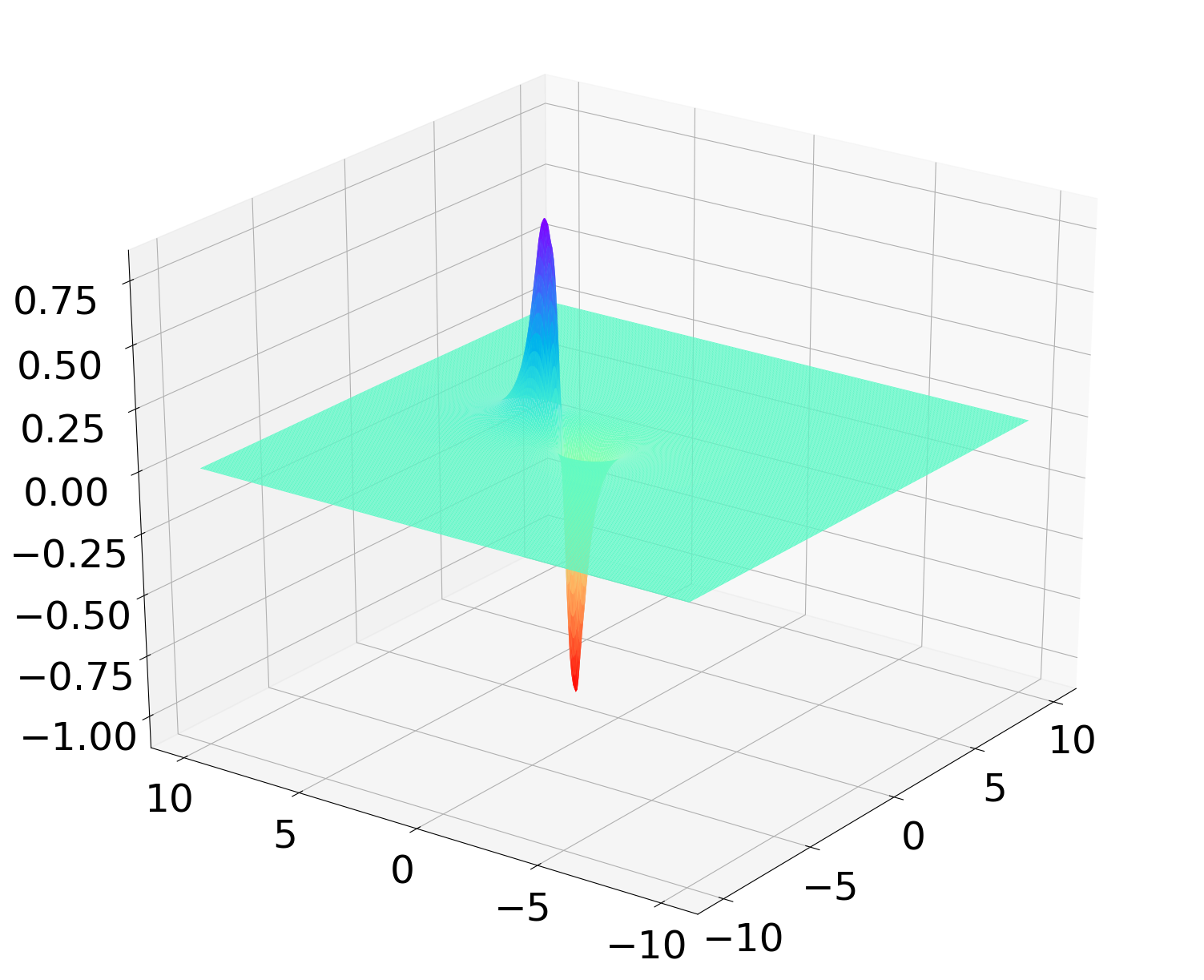}
   \put(-220,75){${\bf k}_y$} 
   \put(-40,15){$x$}
   \put(-150,5){$y$}  
  }
  \caption{A distorted Kerr black hole. The form of the constrained fields 
  $(\widehat{N}, {\bf K}, {\bf k}_x, {\bf k}_y)$ is depicted at $z=0.5$ 
  resulting from the numerical solution of the constraints 
  \eqref{hamiltonian_cart}-\eqref{momentum_K_cart} for the distorted Kerr 
  black hole with input parameters   $(M,\,d,\, \upsilon,\,a)=(0.5,\,2.5,\, 
  0.6,\,0.15)$ and $a'=0.45$ discussed in the present section.}
 \label{Fig:Kerr_dist}
\end{figure}
These are initial data for a dynamical space-time corresponding to a Kerr 
black hole plus non-trivial gravitational radiation \cite{Alcubierre2008}.
This kind of non-stationary radiative Kerr black holes can be generated 
by just altering the Kerr-Schild data used to prescribe initial and boundary 
conditions to the constrained fields from the Kerr-Schild data used to 
specify the values of the freely-specifiable fields throughout $\Sigma$. 

In principle, any modification of the parameters $M, d, \upsilon, a$ of 
the Kerr-Schild data can lead to distorted black hole configurations. 
Here, we choose to alter only the spin. Therefore, we choose the spin of 
the Kerr-Schild data used to prescribe the initial and boundary values 
of $(\widehat N,\, {\bf k}_A,\, {\bf K})$ to be three times bigger than 
the spin of the Kerr-Schild data used to specify the values of ($\widehat 
N^A,\, \widehat \gamma_{AB},\, \boldsymbol\kappa,\, \overset{\circ}{\bf 
K}_{AB})$ throughout $\Sigma$, i.e. $a' = 3\, a$. Our numerical setting 
is similar to the one depicted on the left panel of Fig.~\ref{Fig:Kerr_exact}, 
namely a black hole of mass $M = 0.5$ is placed on the $y$-axis at distance 
$d = 2.5$ from the origin, moves along the $x$-axis with velocity $\upsilon 
= 0.6$, and carries spin $a = 0.15$ along the $z$-axis. The computational 
domain is the same as in the stationary case, i.e. $-4 \leq x \leq 4$,\, 
$-1.5 \leq y \leq 6.5$, and $0 \leq z \leq 4$. 
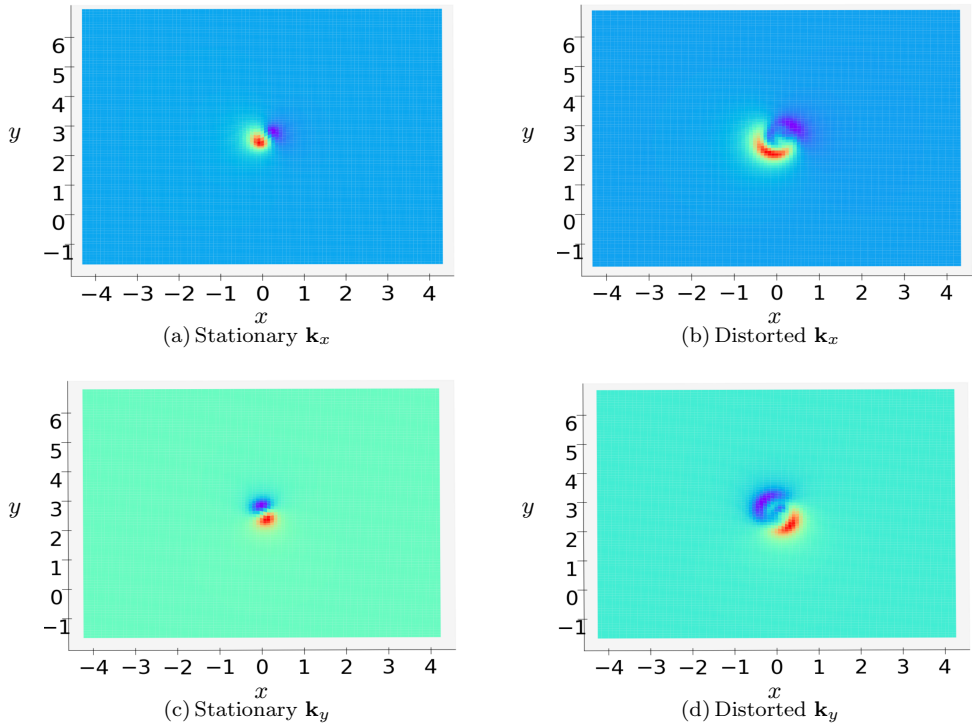
\begin{figure}[htb]
 \centering 
  \subfigure[\,Stationary ${\bf k}_x$]{
   \includegraphics[height=4cm,width=5.5cm]{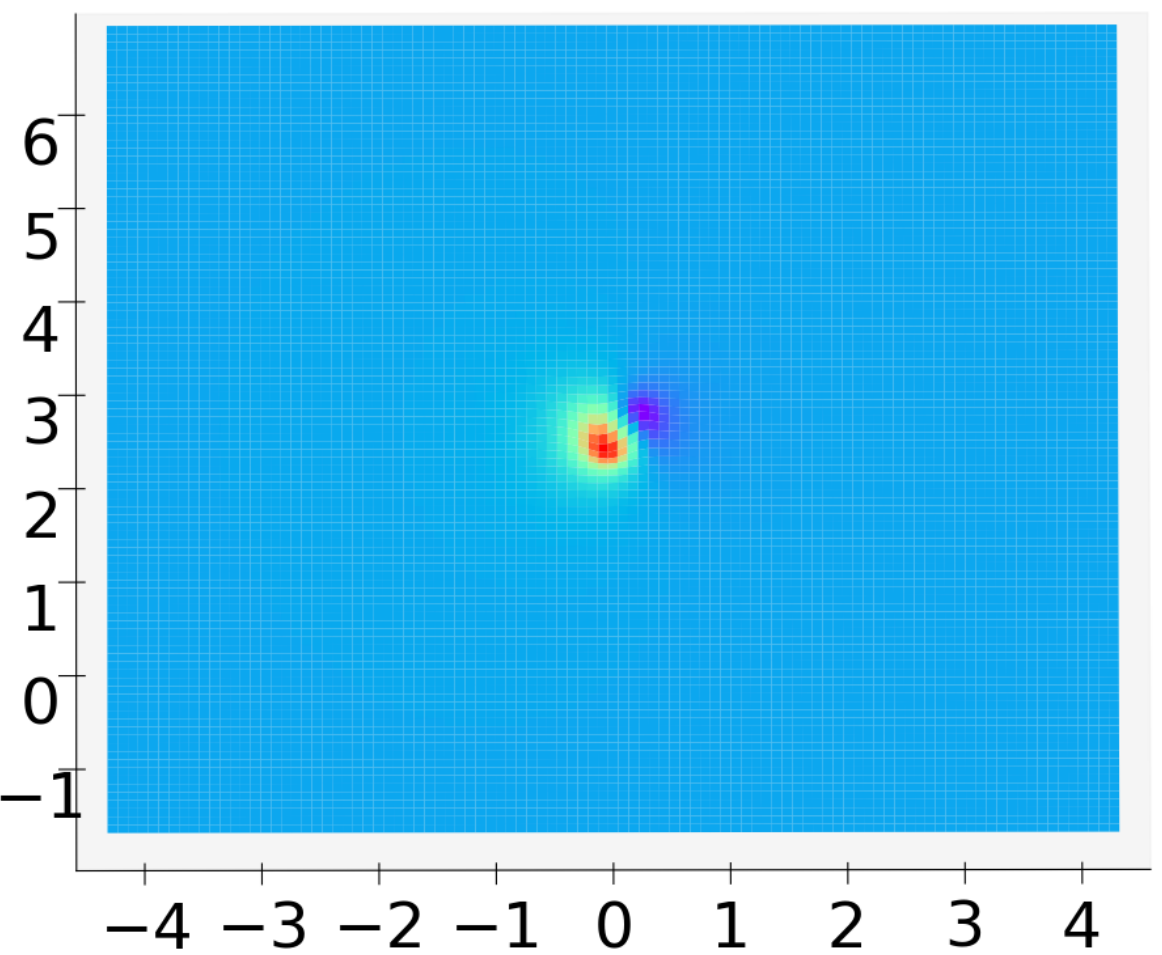} 
  \put(-77,-8){$x$}
  \put(-170,62){$y$} 
  } \hspace{1cm}
  \subfigure[\,Distorted ${\bf k}_x$]{
   \includegraphics[height=4cm,width=5.5cm]{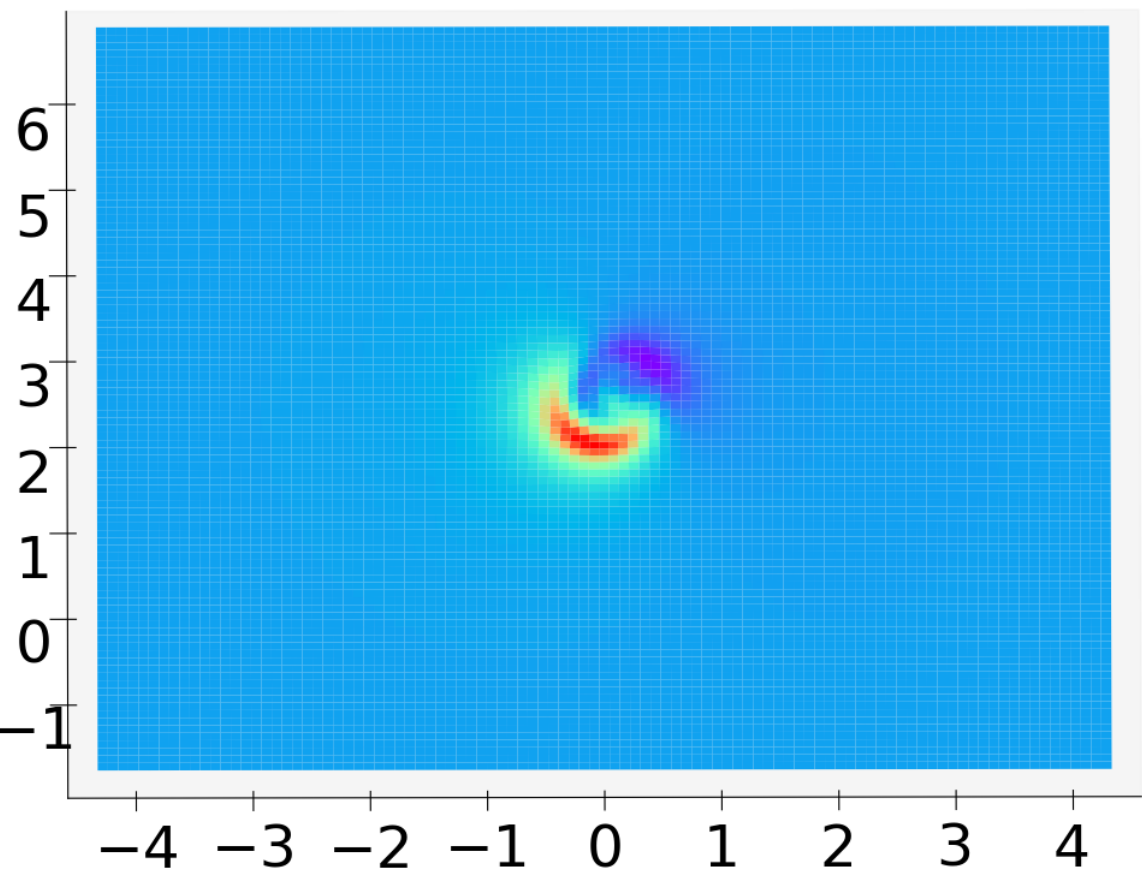} 
  \put(-77,-8){$x$}
  \put(-170,62){$y$} 
  }
  \subfigure[\,Stationary ${\bf k}_y$]{
   \includegraphics[height=4cm,width=5.5cm]{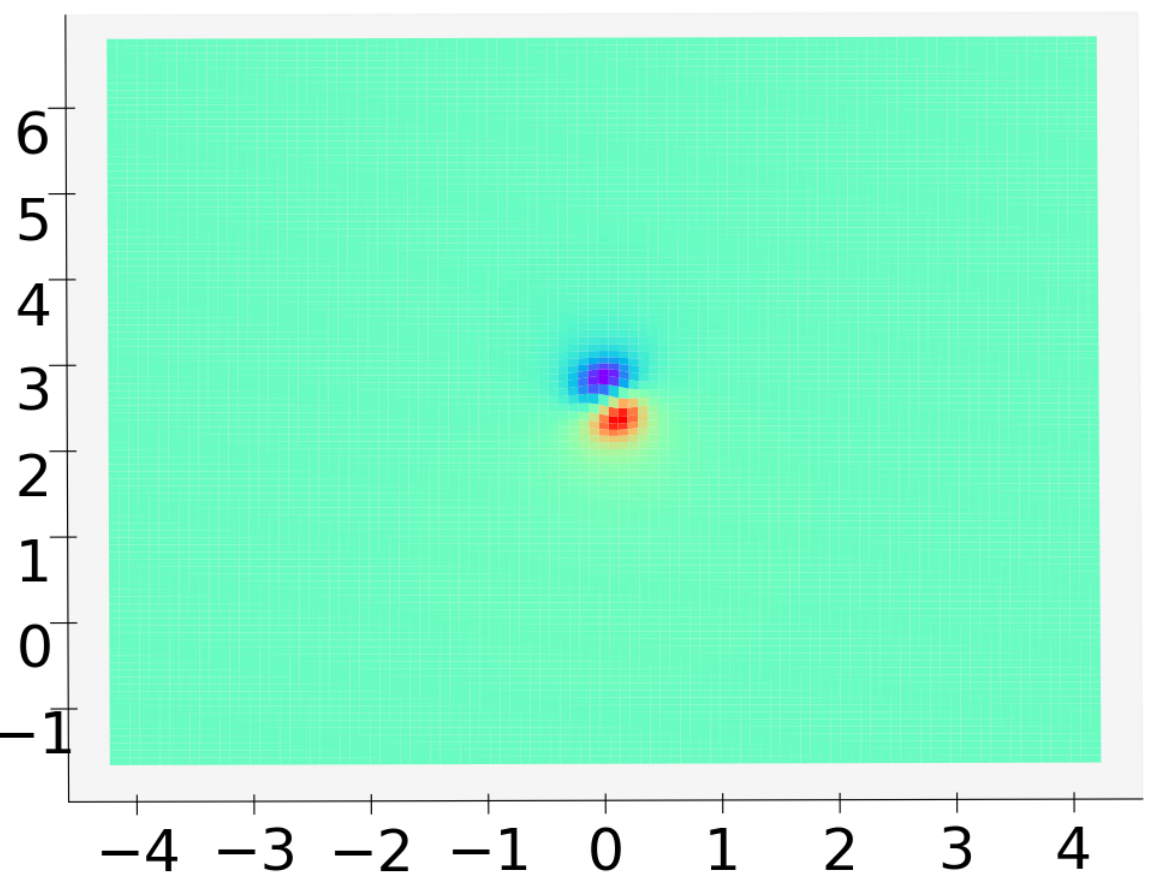} \hspace{1cm} 
  \put(-77,-8){$x$}
  \put(-170,62){$y$} 
  } \hspace{1cm}
  \subfigure[\,Distorted ${\bf k}_y$]{
   \includegraphics[height=4cm,width=5.5cm]{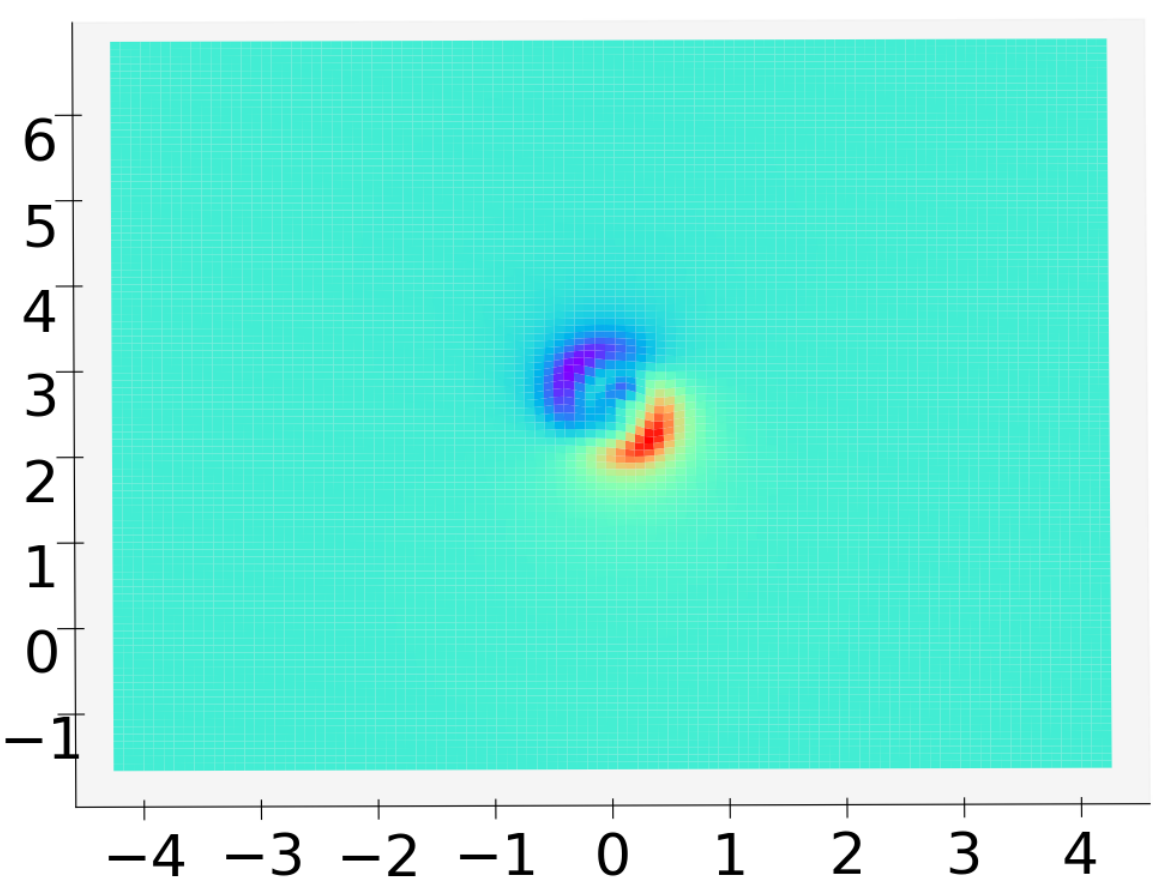} 
  \put(-77,-8){$x$}
  \put(-170,62){$y$}  
  }
  \caption{Comparison of the stationary and distorted Kerr black holes 
  considered in the present work. The left and right panels depict the 
  components of the constraint field ${\bf k}_A$ for the stationary and 
  distorted Kerr black holes studied in Sec.~\ref{sec:single_exact} and 
  the present section, respectively.}
 \label{Fig:Kerr_comparison}
\end{figure}

Fig.~\ref{Fig:Kerr_dist} depicts the constrained fields $(\widehat N,\, 
{\bf k}_A,\, {\bf K})$ resulting from the numerical solution of the system 
\eqref{hamiltonian_cart}-\eqref{momentum_K_cart} for the aforedescribed 
configuration at $z=0.5$. Notice that based on these numerical values 
of $(\widehat N,\, {\bf k}_A,\, {\bf K})$, the physical quantities $(h_{ij}, 
K_{ij})$ can be reconstructed directly from the expressions \eqref{h_K_decomp}. 

The left and right panels of Fig.~\ref{Fig:Kerr_comparison} depict the 
components of the constrained field ${\bf k}_A$ for the stationary and 
distorted Kerr solution considered in Sec.~\ref{sec:single_exact} and 
the present section, respectively. From them it can be safely concluded 
that the distorted solution of Fig.~\ref{Fig:Kerr_dist} departs from the 
stationary solution obtained in Sec.~\ref{sec:single_exact}. The effect 
of using larger spin, i.e. $a'=3a$, to prescribe the initial and boundary 
conditions of $(\widehat N,\, {\bf k}_A,\, {\bf K})$ is clearly visible 
on the graphs of the distorted ${\bf k}_A$, see right panels of 
Fig.~\ref{Fig:Kerr_comparison}. 

\begin{figure}[htb]
 \centering
  \includegraphics[height=5.5cm]{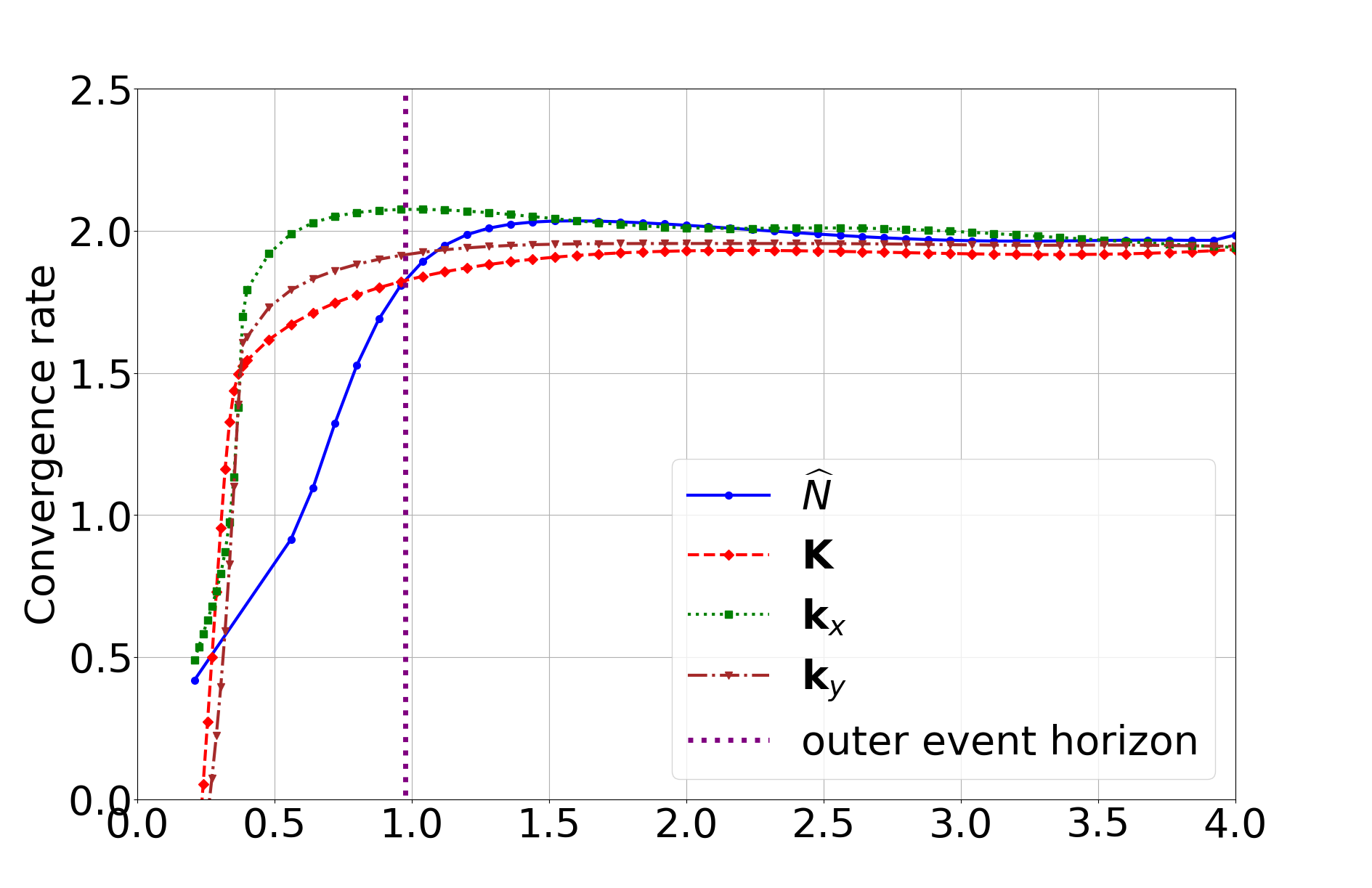}
  \put(-27,0){$z$}
  \caption{Convergence rates of the constraint fields. The convergence 
  of each one of the constrained fields is depicted as a function of the 
  ``temporal" coordinate $z$. The expected behaviour is observed, namely 
  second order convergence in the regions far from the $z=0$ plane and 
  gradual drop of the convergence rates as the $z=0$ plane is approached. 
  The radius of the outer event horizon of the considered distorted Kerr 
  black hole is $r_H \approx 0.977$.}
 \label{Fig:KerrDistConv} 
\end{figure}

We turn now to the convergence analysis of our numerical solution. 
In contrast to Sec.~\ref{sec:exact_sol}, here we do not have an exact 
solution to compare our numerical results with. Therefore, we have 
to proceed in accordance with the discussion below \eqref{conv_rate} 
at the end of Sec.~\ref{sec:numer_scheme}. Accordingly, the dynamical 
behaviour of the convergence rate \eqref{conv_rate} for each of the 
constraints fields is depicted in Fig.~\ref{Fig:KerrDistConv}. It 
is apparent that the convergence rates of all the constrained fields 
are around $2$ for most of the evolution and drop gradually, as expected, 
the closer we get to the $z=0$ plane. 



\subsubsection{Binary black hole systems}
\label{sec:binary_Kerr}

We turn now to the main objective of the present work which is the 
construction of initial data for binary systems of black holes. 
The importance of constructing such kind of data in the study of 
the dynamics of black hole binaries has been already stressed in 
Sec.~\ref{sec:intro}. In the following, we will consider a quite 
general binary black hole configuration consisting of a pair of 
boosted Kerr black holes. As it was already discussed in Sec.~\ref{sec:IBVP}, 
the binary nature of these data dictates the use of the superposed 
Kerr-Schild black hole data \eqref{binaryKS} for prescribing the 
initial and boundary values of the constrained fields $(\widehat 
N,\, {\bf k}_A,\, {\bf K})$ and the values of the freely-specifiable 
fields $(\widehat N^A,\, \widehat \gamma_{AB},\, \boldsymbol\kappa,\, 
\overset{\circ}{\bf K}_{AB})$ throughout $\Sigma$. 
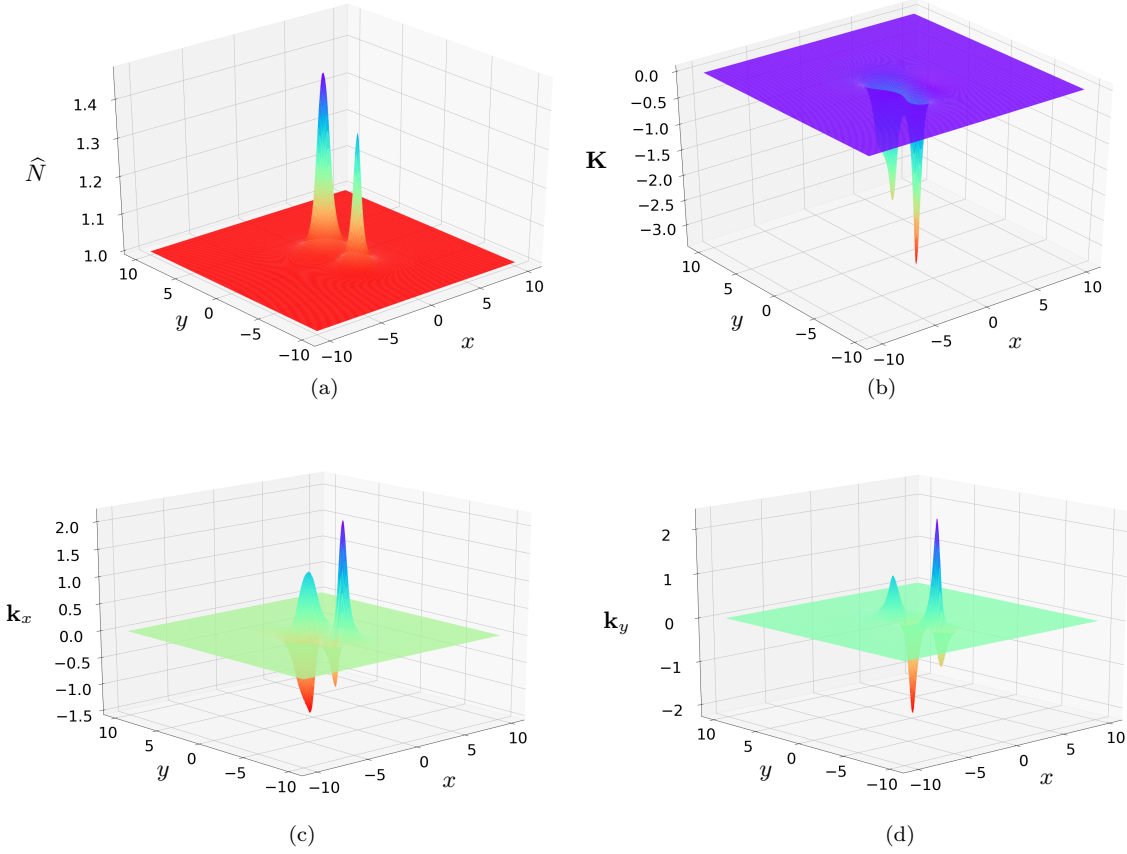
\begin{figure}[htb]
 \centering
  \subfigure[]{
   \includegraphics[scale=0.165]{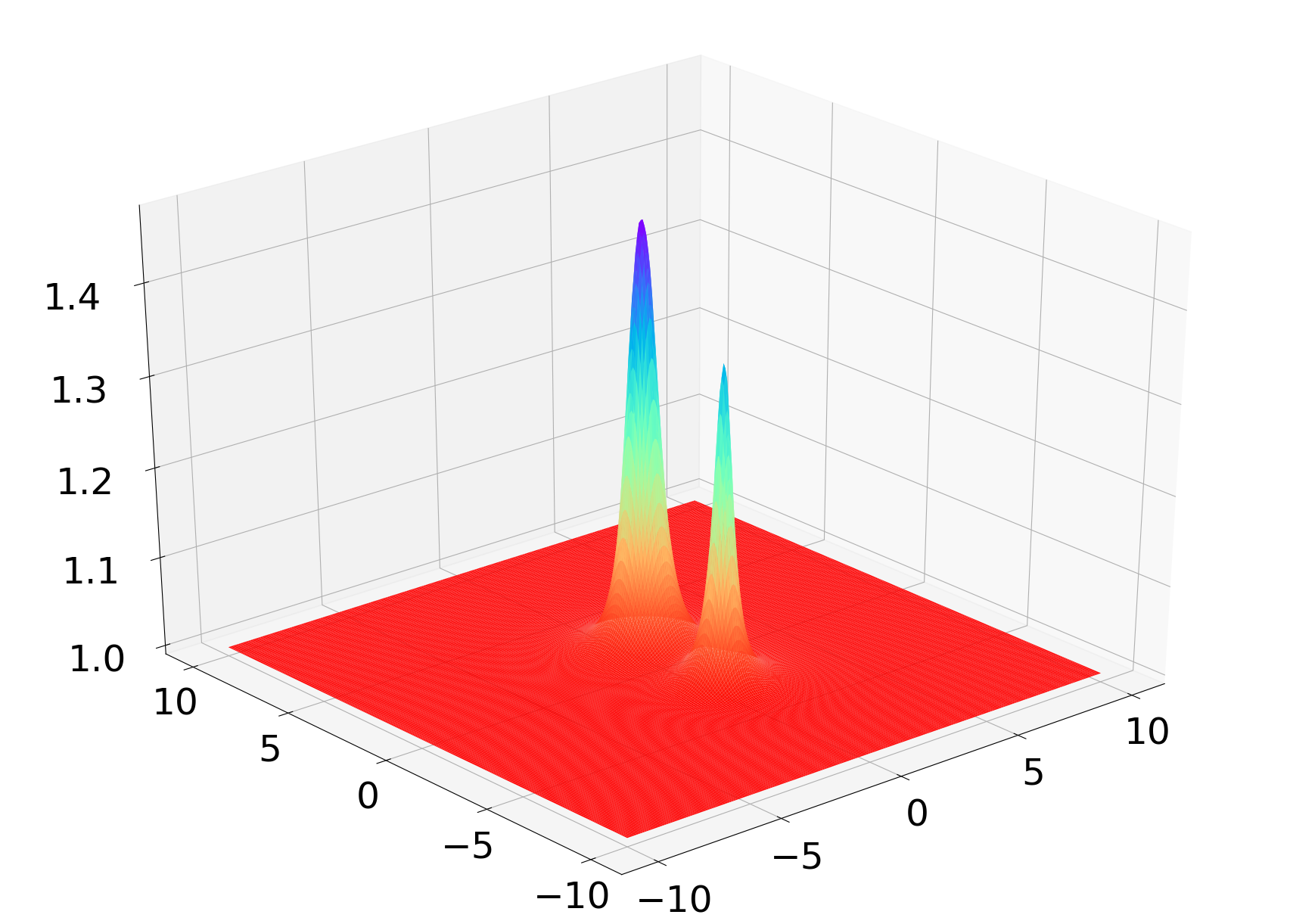}
   \put(-215,70){$\widehat{N}$} 
   \put(-50,7){$x$}
   \put(-158,15){$y$} 
   \label{Fig:binary_lapse}
  } 
  \subfigure[]{
   \includegraphics[scale=0.165]{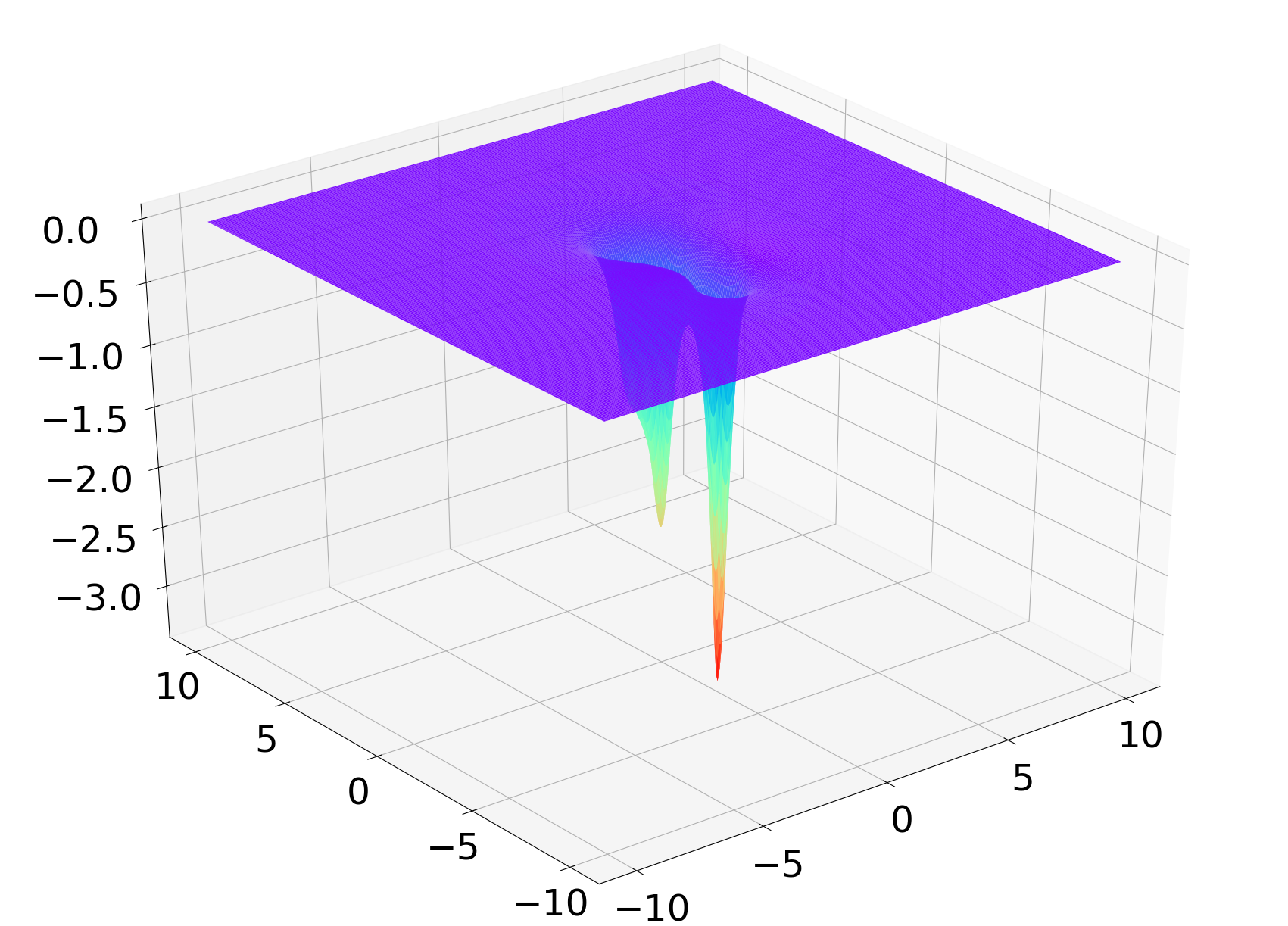}
   \put(-215,75){${\bf K}$} 
    \put(-55,7){$x$}
   \put(-160,15){$y$}   
  }
  \subfigure[]{
   \includegraphics[scale=0.165]{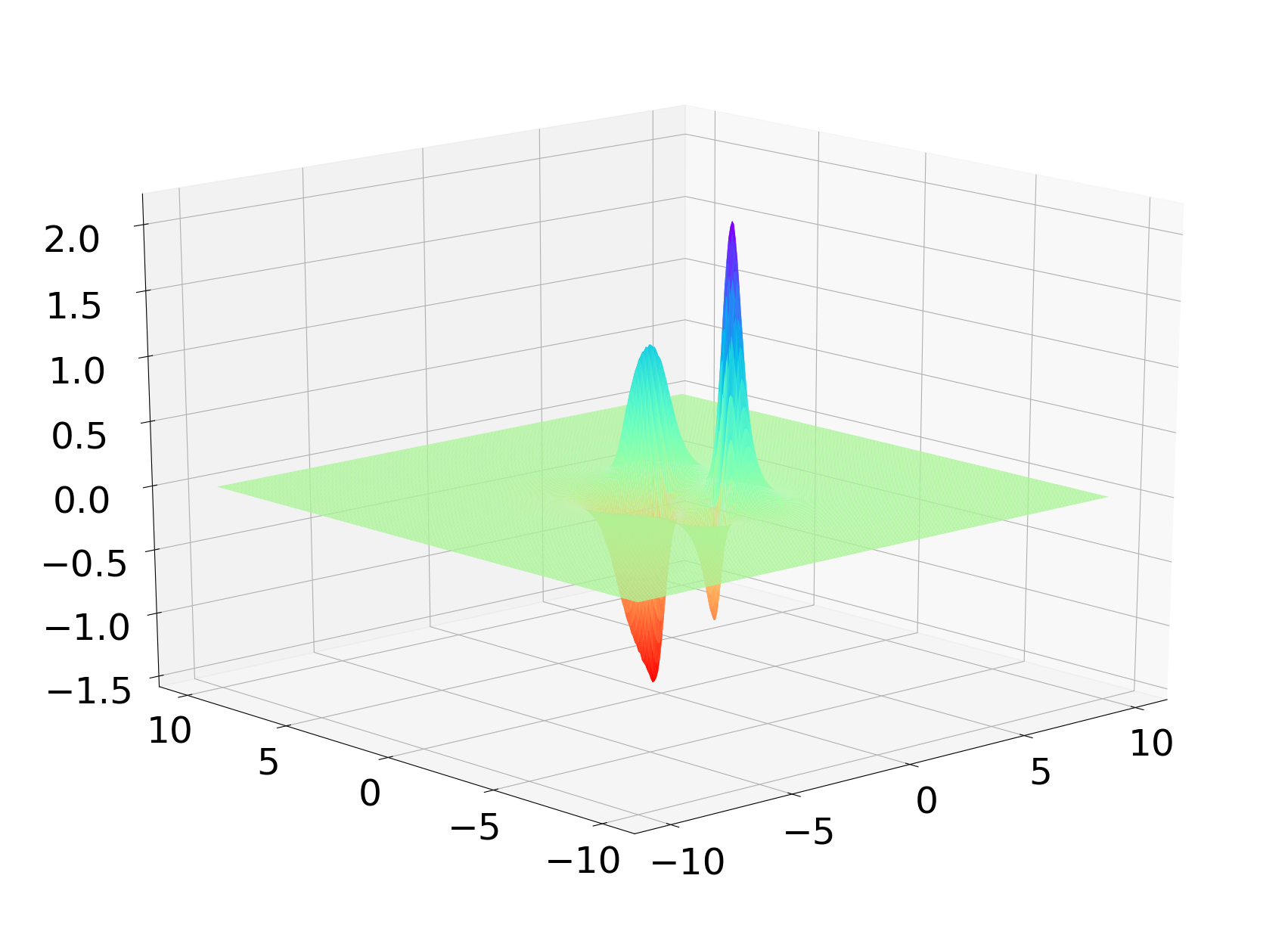}
   \put(-215,73){${\bf k}_x$}
   \put(-50,10){$x$}
   \put(-158,14){$y$}   
  } \hspace{0.5cm}
  \subfigure[]{
   \includegraphics[scale=0.165]{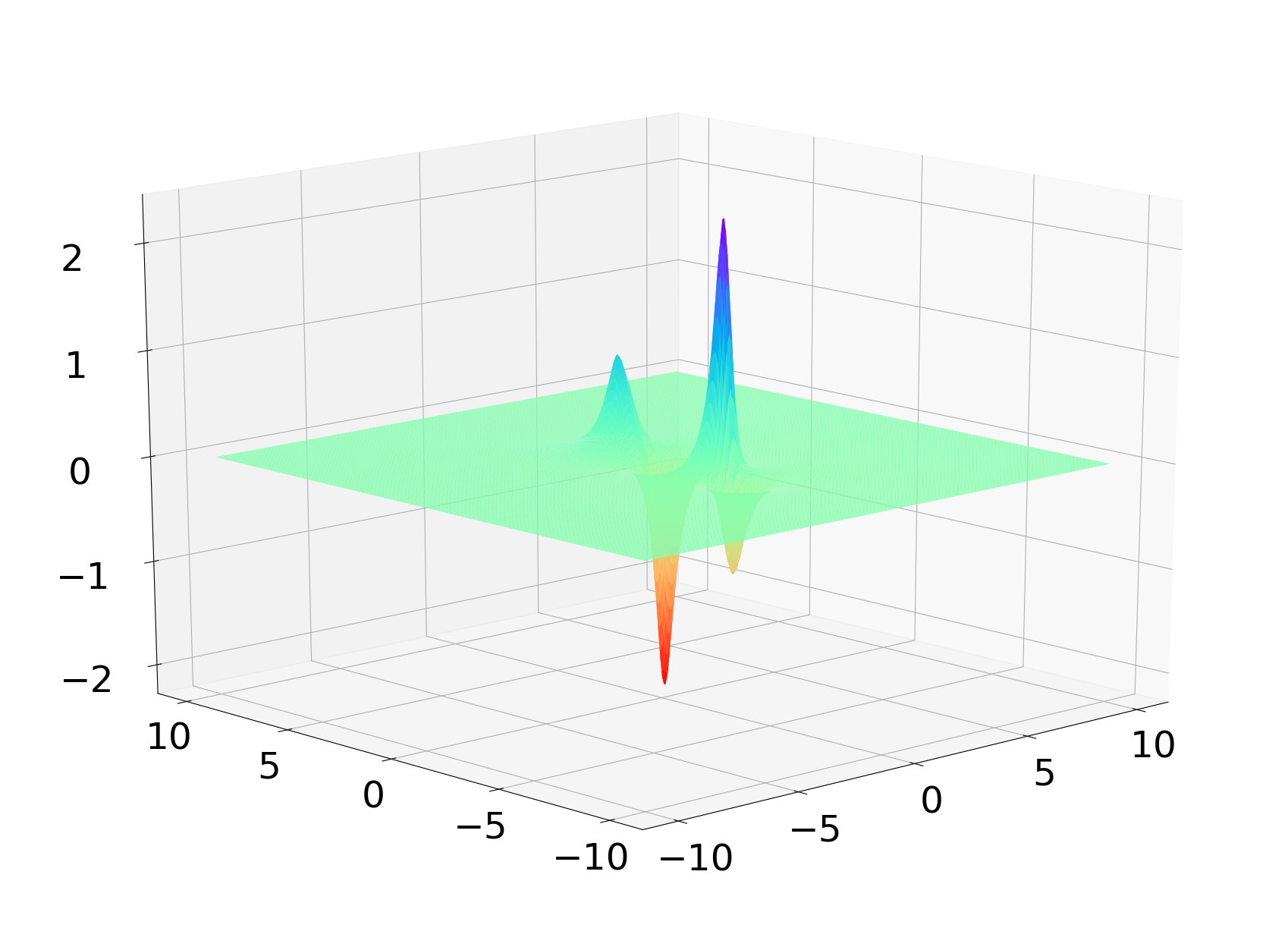}
   \put(-215,70){${\bf k}_y$} 
   \put(-50,11){$x$}
   \put(-156,14){$y$}   
  }
  \caption{A binary system of two Kerr black holes. The numerically 
  computed constrained fields $(\widehat{N},{\bf K}, {\bf k}_x,{\bf k}_y)$ 
  are depicted at $z=0.5$ for the binary configuration with input 
  parameters $(M^{[1]},\, d^{[1]},\, \upsilon^{[1]},\, a^{[1]})= 
  (0.72,\, 1,\, 0.1,\, 0.6)$ and $(M^{[2]},\, d^{[2]},\, \upsilon^{[2]},\, 
  a^{[2]})=(0.25,\, -3,\, -0.3,\, -0.2)$.}
 \label{Fig:binary}
\end{figure}

Our numerical setup consists of two Kerr black holes with masses 
$M^{[1]} = 0.72$ and $M^{[2]}=0.25$ located on the $y$-axis 
at distance $d^{[1]}=1$ and $d^{[2]}=-3$, respectively, from the 
origin. The black holes are confined to the $z=0$ plain, move along 
the $x$-axis with velocities $\upsilon^{[1]}=0.1$ and $\upsilon^{[2]}
=-0.3$, and carry anti-parallel spins $a^{[1]}=0.6$ and $a^{[2]}=
-0.2$ along the $z$-axis. This setting is similar to the one depicted 
in Fig.~\ref{Fig:binaryIBVP}. Note that the input parameters have 
been chosen in such a way that the total ADM centre of mass and linear 
momentum of the binary system are zero, see \cite{Racz2018}. The 
computational domain is $ -10 \leq x \leq 10$, $-10 \leq y \leq 10$ 
and $0 < z \leq 10$. 

\begin{figure}[htb]
 \centering
  \subfigure[\,General case.]{
   \includegraphics[scale=0.16]{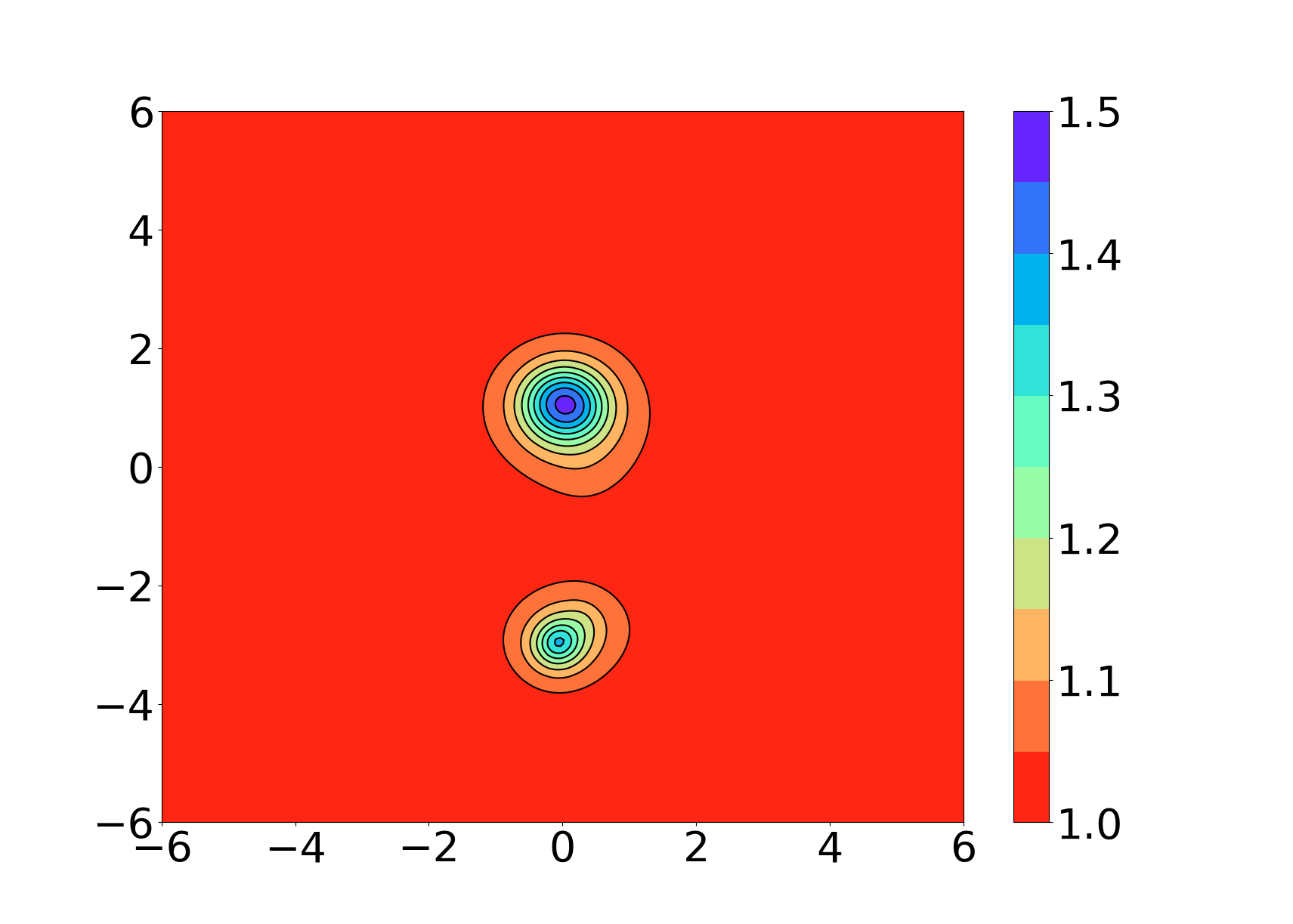}
   \put(-114,0){$x$}
   \put(-195,68){$y$} 
   \label{Fig:binary_contour1}
  } 
  \subfigure[\,$\upsilon^{[1]}=\upsilon^{[2]}=0$.]{
   \includegraphics[scale=0.16]{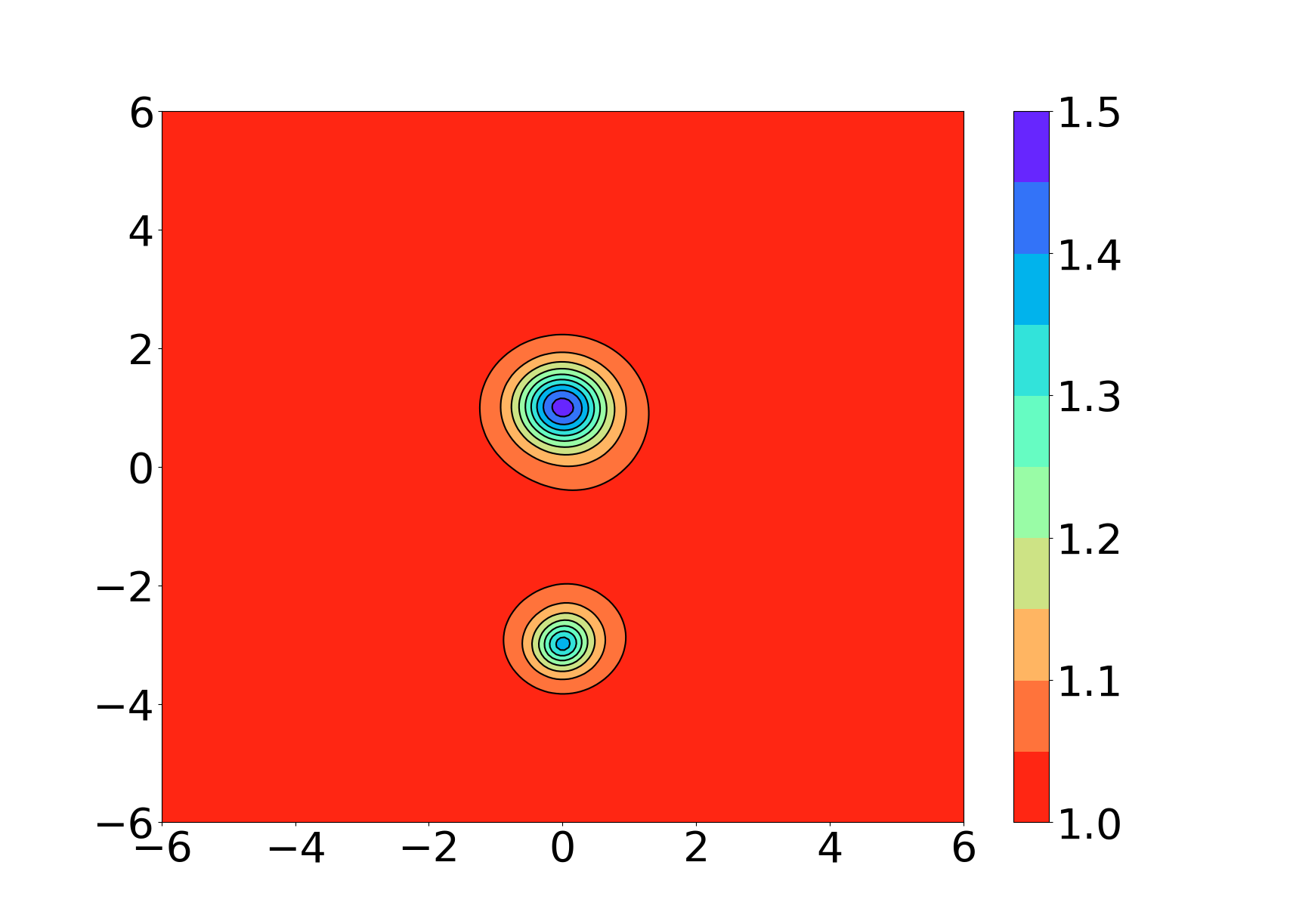}
   \put(-114,0){$x$}
   \put(-195,68){$y$}
   \label{Fig:binary_contour2}  
  } \\ \vspace{-0.3cm}
  \subfigure[\,$a^{[1]}=a^{[2]}=0=\upsilon^{[1]}=\upsilon^{[2]}$.]{
   \includegraphics[scale=0.16]{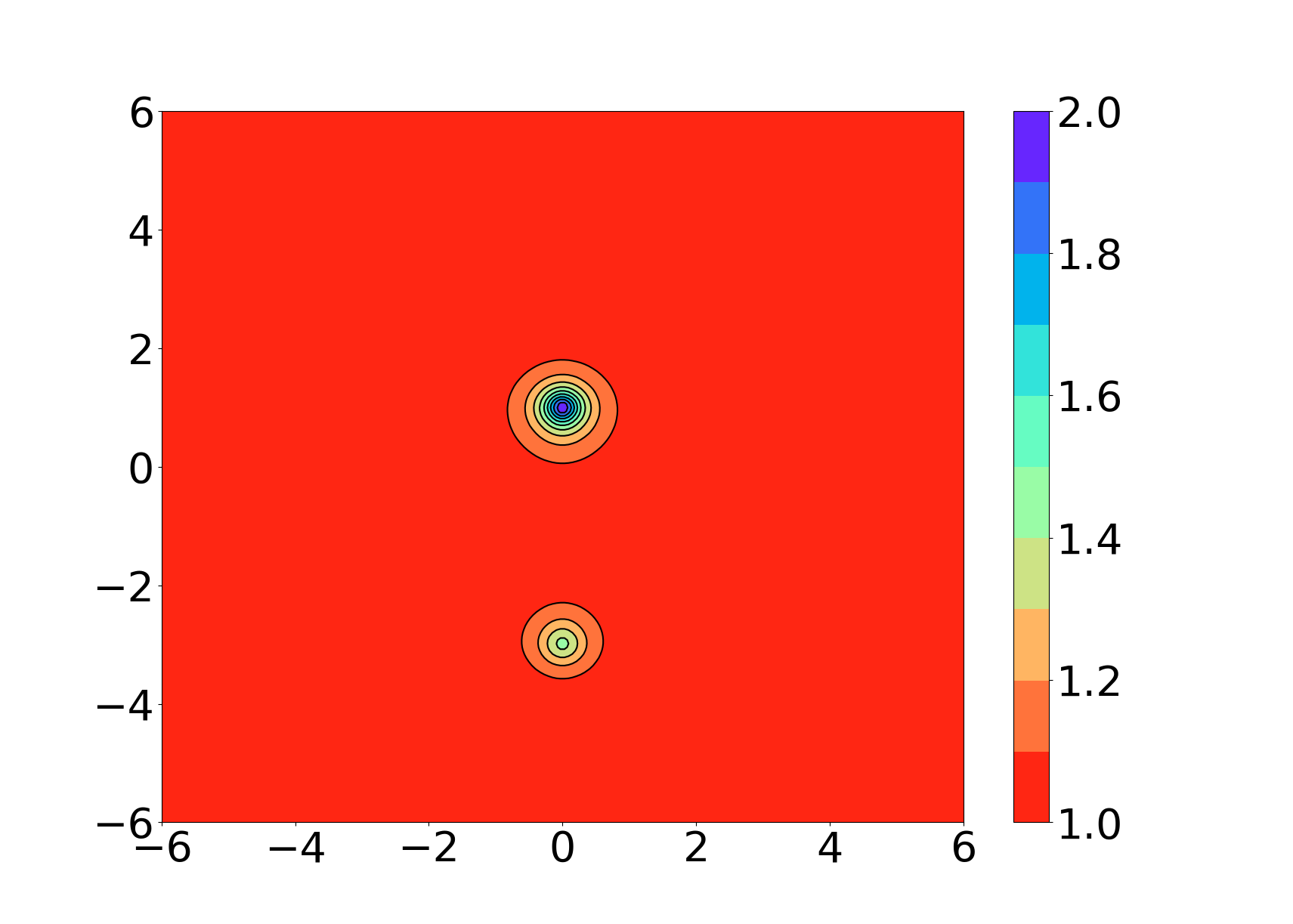}
   \put(-114,0){$x$}
   \put(-195,68){$y$}
   \label{Fig:binary_contour3}  
  } 
  \subfigure[\,$d'^{[1]}<d^{[1]}$, $d'^{[2]}<d^{[2]}$.]{
   \includegraphics[scale=0.16]{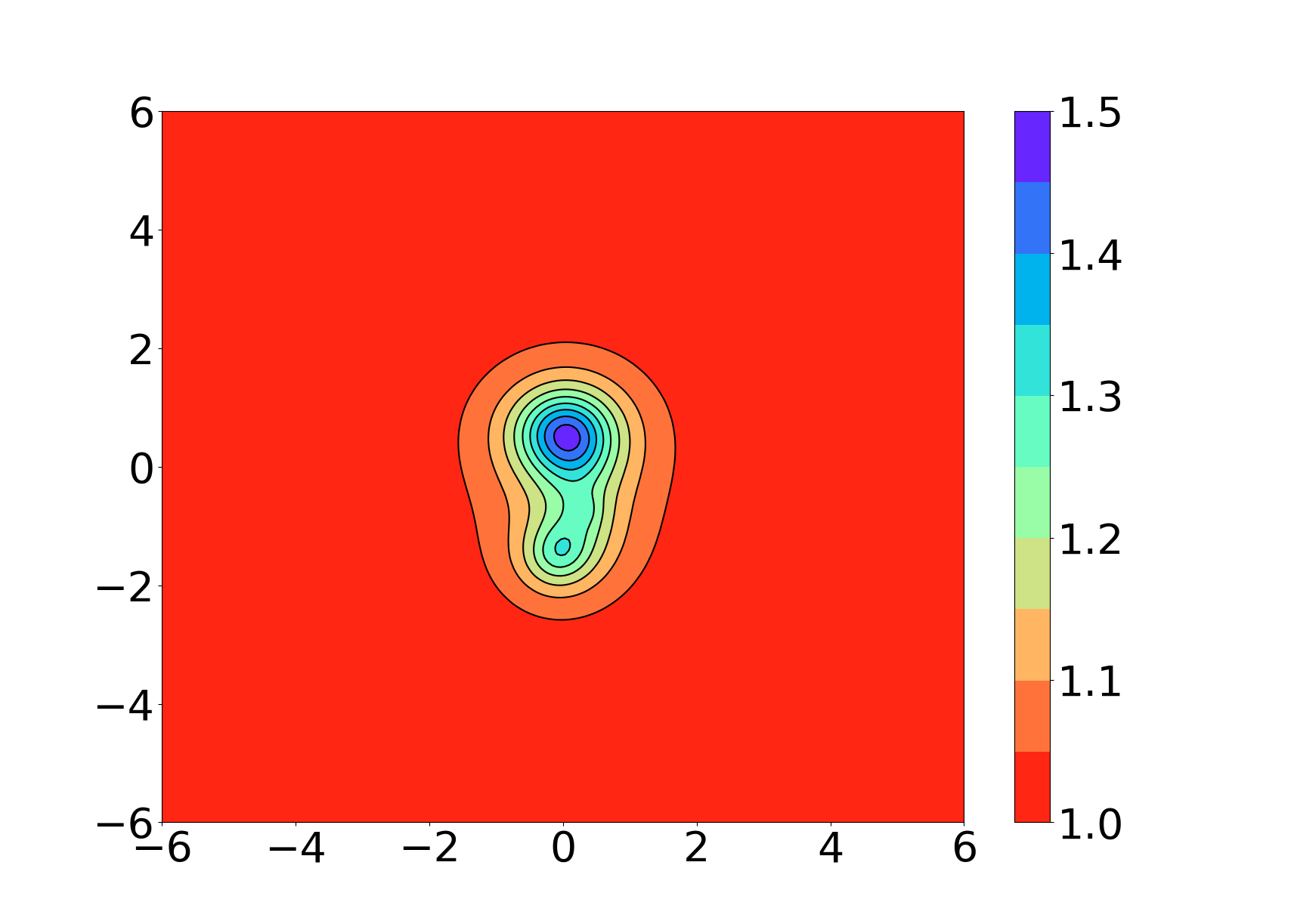}
   \put(-114,0){$x$}
   \put(-195,68){$y$}
   \label{Fig:binary_contour4} 
  }
  \caption{Contour plots of the field $\widehat{N}$ for different 
  values of the input parameters of the binary system considered 
  in the present section. Notice that the expected behaviour of 
  the field $\widehat{N}$, under the considered changes of the 
  input parameters, is successfully reproduced by our code.}
 \label{Fig:binary_behaviour}
\end{figure}
The form of the numerically computed constrained fields $(\widehat{N}, 
{\bf K}, {\bf k}_x,{\bf k}_y)$ resulting from the solution of the 
constraints \eqref{hamiltonian_cart}-\eqref{momentum_K_cart} at 
$z=0.5$ is depicted in Fig.~\ref{Fig:binary}. Notice that for the 
above choice of black hole masses and spins, the radii of their 
respective outer event horizons are equal to $r_H^{[1]} \approx 
1.115$ and $r_H^{[2]}=0.4$. Therefore, the $z=0.5$ plain, on which 
the constrained fields of Fig.~\ref{Fig:binary} have been computed, 
is well inside the outer event horizon of the more massive black hole but 
still outside the outer event horizon of the less massive one. The binary 
nature of the constructed initial data is clearly visible on the plots 
of the four constrained fields---compare with Fig.~\ref{Fig:Kerr_dist}. 

As a first test of our implicit numerical scheme, we will study its 
behaviour with the change of the input parameters of the considered 
binary system. As reference we will use the contour plot of the field 
$\widehat{N}$ at $z=0.5$ depicted in Fig.~\ref{Fig:binary_contour1}. 
Observe that the contours of the two black holes on this plot are 
slightly deformed because of their non-vanishing linear velocity. 
Intuitively, one would expect that when the linear velocities of the 
black holes vanish, then their  contours will acquire a more symmetric 
shape. Clearly, our numerical findings in the case $\upsilon^{[i]}=0$, 
depicted in Fig.~\ref{Fig:binary_contour2}, fulfil this expectation. 
If in addition the spins of the black holes are also set to zero then, 
as shown in Fig.~\ref{Fig:binary_contour3}, the symmetric shape of the 
contours is still preserved but their size has decreased significantly 
because of the lack of spin. Moreover, it is expected that when the 
distance $d=|d^{[1]}|+|d^{[2]}|$ between the black holes is gradually 
decreased, then for a specific value of $d$ a third horizon surrounding 
both black holes will form. The contour plot of Fig.~\ref{Fig:binary_contour4}, 
illustrating our numerical findings for the configuration $d\,'<d$, 
points to this direction as thereon a new contour surrounding both 
black holes is clearly visible. 

\begin{figure}[htb]
 \centering
  \includegraphics[height=5.5cm]{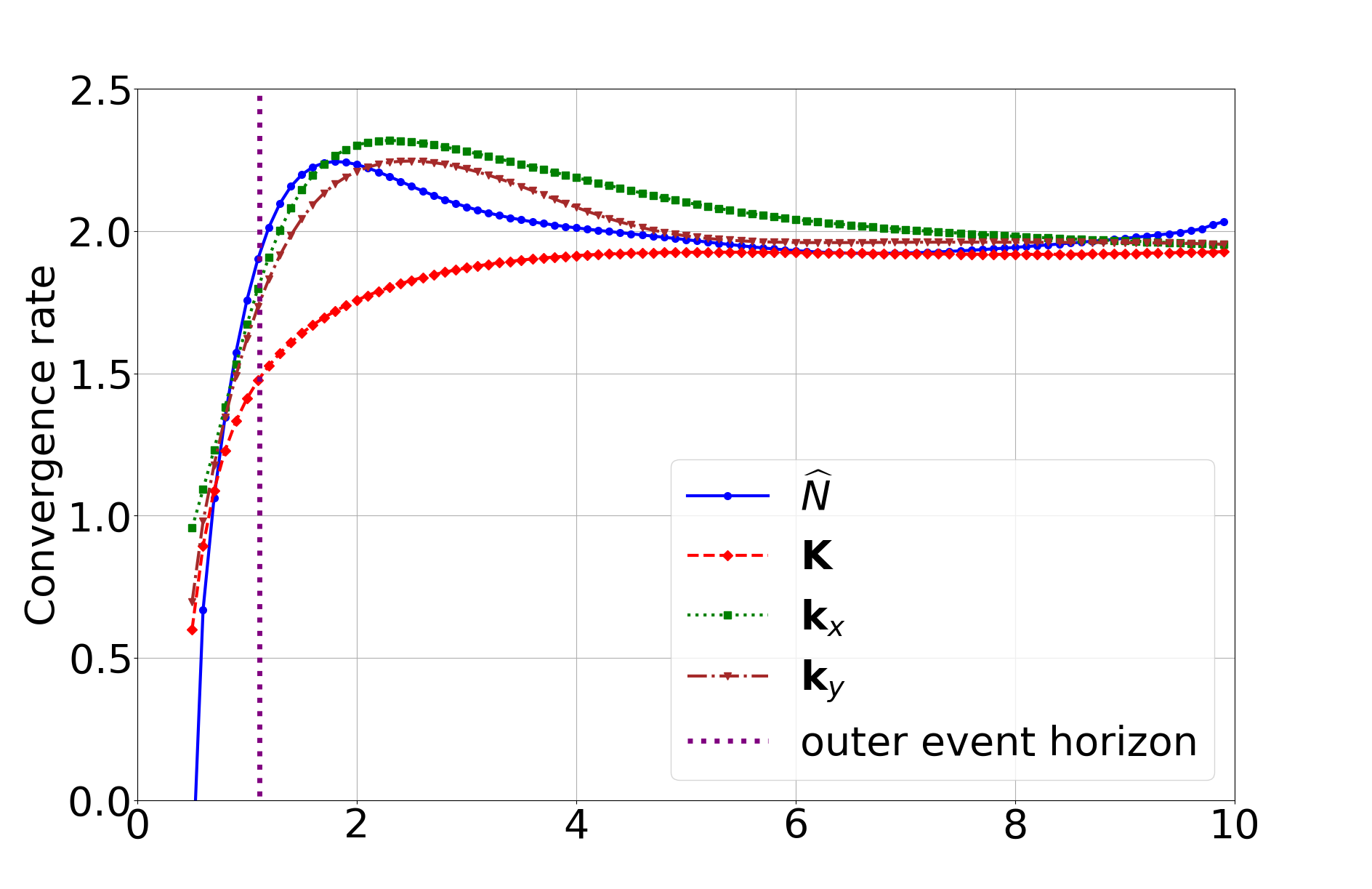}
  \put(-28,0){$z$}
  \caption{Convergence of the constrained fields. The $z$-dependence 
  of the convergence rate for each one of the constrained fields of 
  the numerical solution depicted in Fig.~\ref{Fig:binary} is shown. 
  The convergence is second order and drops progressively while approaching 
  the $z=0$ plane containing the black hole singularities. Notice that 
  the dashed line depicts the location of the outer event horizon of 
  the more massive black hole with radius $r_H \approx 1.115$.}
 \label{Fig:binaryConv} 
\end{figure}

We move on now to the numerical analysis of the solution of 
Fig.~\ref{Fig:binary}. Its basic features are encapsulated in 
Fig.~\ref{Fig:binaryConv}. Therein the dynamical behaviour of the 
convergence rate \eqref{conv_rate} for each one of the constrained 
fields is depicted. As expected the convergence rates of all the 
fields are around $2$ for most of the $z$-evolution and drop gradually 
as we approach the $z=0$ plain where the black hole singularities 
are located. 


\section{Discussion}
\label{sec:discussion}

The purpose of the present paper is to develop an implicit numerical 
scheme for the parabolic-hyperbolic formulation of the constraints 
that can be used to construct highly accurate initial data for single 
and binary black hole configurations. 

The above has been achieved by combining the parabolic-hyperbolic 
formulation of the constraints \cite{Racz2015} with the superposed 
Kerr-Schild black hole type data in the way proposed in \cite{Racz2018,Racz2018-SM}. 
In this setting, the constraints are formulated as an initial-boundary 
value problem for the fields $(\widehat N, {\bf k}_{i}, {\bf K})$ on 
an initial data three-surface represented as a cube $\Sigma$ of finite 
side, see Sec~\ref{sec:IBVP}. The resulting constraint equations 
\eqref{hamiltonian_cart}-\eqref{momentum_K_cart} are solved on $\Sigma$ 
given that the values of the freely-specifiable fields $(\widehat N^i, 
\widehat \gamma_{ij}, \boldsymbol\kappa, \interior{\bf K}_{ij})$ and 
of the constrained fields $(\widehat N, {\bf k}_{i}, {\bf K})$ have 
been appropriately provided throughout and on the sides of $\Sigma$, 
respectively. Kerr-Schild data, briefly discussed in Sec.~\ref{sec:IBVP}, 
are then used to prescribe the initial and boundary values of $(\widehat 
N, {\bf k}_{i}, {\bf K})$ on the sides of $\Sigma$ and the values of 
$(\widehat N^i, \widehat \gamma_{ij}, \boldsymbol\kappa, \interior{\bf 
K}_{ij})$ throughout $\Sigma$. 

To solve numerically the constraints in their parabolic-hyperbolic form 
\eqref{hamiltonian_cart}-\eqref{momentum_K_cart}, we chose to use the 
A-stable unconditionally stable alternating direction implicit method 
described in Sec.~\ref{sec:numer_scheme}. The extensive convergence 
analysis of Sec.~\ref{sec:exact_sol} demonstrates the stability, accuracy 
and convergence properties of our numerical scheme. The performance of 
the developed numerical scheme was extensively tested against known 
exact black hole solutions in the single (Schwarzschild and Kerr) and 
binary (Brill-Lindquist) case. Our results, both for single and binary 
black hole configurations, confirm the expected convergence and stability 
features of the implicit method used.

In Sec.~\ref{sec:results} our implicit scheme was used to generate 
initial data for dynamical (single and binary) black hole configurations 
carrying non-trivial gravitational radiation. In the single black 
hole case, initial data for a distorted Kerr black hole were successfully 
constructed in Sec.~\ref{sec:dist_Kerr}. The form of the resulting 
constrained fields is depicted in Fig.~\ref{Fig:Kerr_dist}. This 
result is the first manifestation of the control we have over the 
constructed initial data within R\'azc's method. In contrast to the 
other existing methods, we have full control over the gravitational 
radiation content of the constructed non-stationary initial data as 
its appearance is a result of our conscious choice to alter the 
spin of the used Kerr-Schild data. In the binary black hole case, 
initial data for a binary system of Kerr black holes were constructed 
in Sec.~\ref{sec:binary_Kerr}. This is a first example of the kind 
of binary initial data that can be constructed with the proposed 
implicit numerical scheme. Notice that in order to generate the 
above data sets no boundary conditions were used in the strong 
field regime. 

The above numerical results demonstrate not only the simplicity of 
the parabolic-hyperbolic method but also its all inclusive nature 
as different single and binary black hole configurations can be 
treated within the same numerical setup. Specifically, the only 
parameters that have to be changed to get the different black hole 
initial data constructed in the present work are the input parameters 
characterising the black holes, i.e. $M_i, d_i, \upsilon_i, a_i$. 

As demonstrated in Fig.~\ref{Fig:IBVP}, the lack of any boundary 
conditions in the strong field regime makes the parabolic-hyperbolic 
method highly attractive as the "junk radiation" common to all the 
existing formulations of the constraints could be reduced significantly 
or even entirely suppressed. 

Special attention must be given to the relations \eqref{h_K_decomp} 
as they give us direct control of the physical quantities $(h_{ij}, 
K_{ij})$. Notice that the lack of any conformal rescalings in our 
formulation provides a direct way of reconstructing the original 
fields $(h_{ij}, K_{ij})$ from the ones computed numerically.  

In the context of the present implicit approach, a different 
formulation of the parabolic-hyperbolic system that is based on 
the deviation from a known Kerr-Schild black hole solution is 
currently developed. As was already shown in \cite{Nakonieczna2017}, 
this approach will not only increase the accuracy of the produced 
data and improve the convergence properties of our numerical scheme close 
to the $z=0$ plane, but will also reduce the computational complexity 
of the proposed implicit scheme and the required computational 
resources. In addition, we plan to evolve the constructed initial 
data in collaboration with other numerical relativity groups that 
have already fully developed evolutionary codes. The evolution 
of our data will decidedly answer, among others, the question of 
whether the constructed initial data are ``junk radiation"-free.


\acknowledgments

The author is deeply indebted to Istv\'an R\'acz for introducing 
him to the current topic and for his support and encouragement 
during the course of the present work. This work was supported 
by the POLONEZ programme of the National Science Centre of Poland 
which has received funding from the European Union's Horizon 2020 
research and innovation programme under the Marie Sk{\l}odowska-Curie 
grant agreement No.~665778 and by a STSM Grant from COST Action 
CA16104: Gravitational waves, black holes and fundamental physics 
(GWverse). 


\appendix

\section{Coefficients of the constraint equations}
\label{sec:appendixA}

\noindent Here, we present the full form of the coefficients entering the 
constraints \eqref{hamiltonian_cart}-\eqref{momentum_K_cart}. 

\noindent Coefficients of the parabolic equation \eqref{hamiltonian_cart}:
\begin{align*}
 &A_1 = \widehat{N}^2\, \instar{K}{}^{-1}\, \gamma^{xx}, \qquad
 A_2 = \widehat{N}^2\, \instar{K}{}^{-1}\, \gamma^{yy}, \qquad
 A_3 = 2\, \widehat{N}^2\, \instar{K}{}^{-1}\, \gamma^{xy},  \\
 &A_4 =  \widehat{N}^x - \widehat{N}^2\, \instar{K}{}^{-1}\, \gamma^{AB}\, 
 \widehat{\Gamma}\,^x_{BA}, \qquad
 A_5 =  \widehat{N}^y - \widehat{N}^2\, \instar{K}{}^{-1}\, \gamma^{AB}\, 
 \widehat{\Gamma}\,^y_{BA}.
\end{align*}

\noindent Coefficients of the hyperbolic equation \eqref{momentum_kx_cart}: 
\begin{align*}
 &\qquad\qquad\qquad B_1 = \widehat{N}^x, \qquad  B_2 = \widehat{N}^y, \qquad  
 B_3 = \frac{\widehat{N}}{2}, \\
 B_4 &= \left(\partial_x \widehat{N}^x - \instar{K} \right) {\bf k}_x - 
 \widehat{N}\, \gamma^{CB} \left(\partial_B \overset{\circ}{\bf K}_{Cx} - 
 \widehat{\Gamma}\,^D_{BC}\, \overset{\circ}{\bf K}_{Dx} - \widehat{\Gamma}\,^D_{Bx}\, 
 \overset{\circ}{\bf K}_{CD} \right) +\\
 &+ {\bf k}_y\, \partial_x \widehat{N}^y + \widehat{N}\, \partial_x \boldsymbol\kappa - 
 \widehat{N} \left(\boldsymbol\kappa - \frac12\, {\bf K} \right) \dot{\widehat{n}}_x + 
 \widehat{N}\, \dot{\widehat{n}}{}^B\, \overset{\circ}{\bf K}_{Bx}.
\end{align*}

\noindent Coefficients of the hyperbolic equation \eqref{momentum_ky_cart}: 
\begin{align*}
 &\qquad\qquad\qquad C_1 = \widehat{N}^x, \qquad  C_2 = \widehat{N}^y, \qquad  
 C_3 = \frac{\widehat{N}}{2}, \\
 C_4 &= \left(\partial_y \widehat{N}^y - \instar{K} \right) {\bf k}_y - 
 \widehat{N}\, \gamma^{CB} \left(\partial_B \overset{\circ}{\bf K}_{Cy} - 
 \widehat{\Gamma}\,^D_{BC}\, \overset{\circ}{\bf K}_{Dy} - \widehat{\Gamma}\,^D_{By}\, 
 \overset{\circ}{\bf K}_{CD} \right) +\\
 &+ {\bf k}_x\, \partial_y \widehat{N}^x + \widehat{N}\, \partial_y \boldsymbol\kappa - 
 \widehat{N} \left(\boldsymbol\kappa - \frac12\, {\bf K} \right) \dot{\widehat{n}}_y + 
 \widehat{N}\, \dot{\widehat{n}}{}^B\, \overset{\circ}{\bf K}_{By}.
\end{align*}

\noindent Coefficients of the hyperbolic equation \eqref{momentum_K_cart} 
\begin{align*}
 &D_1 = \widehat{N}^x, \qquad  D_2 = \widehat{N}^y, \qquad D_3 = \widehat{N}\, 
 \gamma^{xx}\, \partial_x {\bf k}_x, \qquad D_4 = \widehat{N}\, \gamma^{yx}\, 
 \partial_x {\bf k}_y, \\
 &D_5 = \widehat{N}\, \gamma^{xy}\, \partial_y {\bf k}_x, \qquad D_6 = 
 \widehat{N}\, \gamma^{yy}\, \partial_y {\bf k}_y, \\
 &D_7 = -\frac{\instar{K}}{2}\, {\bf K} - \widehat{N}\, \gamma^{AB}\, 
 \widehat{\Gamma}\, ^C_{AB}\, {\bf k}_C - 2\, \widehat{N}\, \dot{\widehat{n}}{}^A\, 
 {\bf k}_A + \instar{K}\, \boldsymbol\kappa - \overset{\circ}{\bf K}_{AB} \, 
 \instar{K}{}^{AB}. 
\end{align*}

All the quantities involved in the above expressions have been defined 
in Sec.~\ref{sec:parab-hyperb} except of the fields $\widehat{\Gamma}$, 
which are the Christoffel symbols associated with the two-metric 
$\widehat{\gamma}_{AB}$.


\bibliography{bibliography}

\end{document}